\documentclass[journal=jpcafh,manuscript=article]{achemso}

\usepackage{chemformula} 
\usepackage[T1]{fontenc} 

\usepackage{amsfonts}
\usepackage{amssymb}
\usepackage{graphicx}
\usepackage{subfigure}
\usepackage{amsmath}
\usepackage[english]{babel}
\usepackage{color}
\usepackage{epstopdf}
\usepackage{footnote}
\usepackage{multirow}

\newcommand{\topcaption}{
\setlength{\abovecaptionskip}{0pt}
\setlength{\belowcaptionskip}{15pt}
\caption}

\author{Kaijun Shen}
\affiliation[First University]{School of Materials Science and Engineering, Nanyang Technological University, Singapore 639798, Singapore}
\author{Kewei Sun}
\affiliation[First University]
{School of Materials Science and Engineering, Nanyang Technological University, Singapore 639798, Singapore}
\alsoaffiliation[Second University]
{School of Science, Hangzhou Dianzi University, Hangzhou 310018, China}
\author{Yang Zhao}
\affiliation[First University]
{School of Materials Science and Engineering, Nanyang Technological University, Singapore 639798, Singapore}
\email{YZhao@ntu.edu.sg}

  \title{Simulation of emission spectra of transition-metal dichalcogenide monolayers with the multimode Brownian oscillator model}

\abbreviations{IR,NMR,UV}

\SectionNumbersOn
\begin{document}

\begin{tocentry}
\centering
\includegraphics[width = 8.25cm, height = 4.45cm]{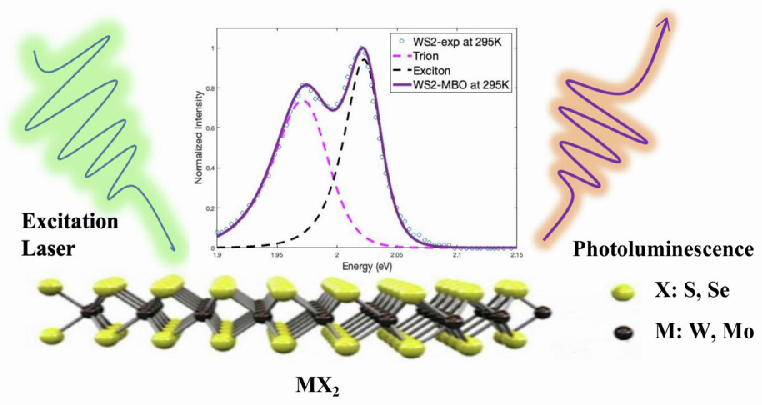}
\end{tocentry}

\begin{abstract}
The multimode Brownian oscillator model is employed to simulate the emission spectra of transition metal dichalcogenide monolayers. Good agreement is obtained between measured and simulated photoluminescence spectra of WSe$_2$,  WS$_2$, MoSe$_2$ and MoS$_2$ at various temperatures.
The Huang-Rhys factor extracted from the model can be associated with that from the modified semi-empirical Varshni equation at high temperatures.
Individual mechanisms leading to the unique temperature-dependent emission spectra of those TMDs are validated by the MBO fitting, while it is in turn  confirmed that the MBO analysis is an effective method for studying the optical properties of TMD monolayers. Parameters extractd from the MBO fitting can be used to explore exciton-photon-phonon dynamics of TMDs in a more comprehensive model.
\end{abstract}


\section{Introduction}
Ultrathin nanosheets of layered transition metal dichalcogenide (TMD) are fundamentally attractive as two-dimensional (2D) semiconducting substitutes to metallic graphene with outstanding properties. For instance, MoS$_2$,  a typical family member of TMDs, possesses high charge mobility~\cite{rad}, strong photoluminescence (PL)~\cite{mak2, spl}, indirect-to-direct band gap transition with direct wide band gap~\cite{yan} (1.2-1.9 eV), and a spin-valley locking relationship~\cite{di}. Similar properties have also been reported for WSe$_2$~\cite{kim}. Moreover, several studies has illustrated the generation and electrostatic manipulation of charged excitons or trions in WSe$_2$~\cite{jones}, MoS$_2$~\cite{mak2} and MoSe$_2$~\cite{ros}, analogous to what was observed earlier in the quantum wells of quasi-2D semiconductors~\cite{fink, hua} . Although these remarkable properties contribute to its application in optoelectronic devices, how to model its absorption and emission spectra and quantify the impact of lattice vibration on the investigated spectra still remains a puzzle. Thus a reasonable and comprehensive physical model is pivotal to figure out the underlying photophysical mechanism.

Here we propose the multimode Brownian oscillator (MBO) model to depict the photophysics mechanism. The MBO model illustrates the electronic relaxation in a two-level system (TLS) accompanied with one (few) primary oscillators(s), which is (are) linearly coupled to a reservoir of secondary modes~\cite{muk}. It therefore absorbs the coupling of nuclear motions to optical transitions for describing the dissipation of thermal bath~\cite{knox, xu}. The crucial parameters in the MBO model are the frequencies of primary oscillators, the damping coefficients and the Huang-Rhys (HR) factors; other influencing parameters involve the 0-0 transition energy and the temperature. The frequency of a primary oscillator adjusts the energy separation between the zero phonon line (ZPL) and its neighboring phonon peak. The damping coefficient represents the coupling strength between the primary oscillator and the bath modes, controlling the smoothness of the PL line shape in addition to broadening the ZPL and phonon side bands (PSBs). The HR factor~\cite{huang} catches the exciton-phonon coupling strength and alters the intensity ratio of one-phonon peak to the ZPL.

The MBO model has been successfully applied to organics systems to analyze the spectra of chromophores in liquid and glasses, such as polyindenofluorenes~\cite{ref14} and pentacene~\cite{ref15}. To our knowledge, MBO model has not been utilized to explain absorption and emission spectra of TMD monolayers. Recently, Qi \emph{et al.}~\cite{Qi} investigated PL spectra of WSe$_2$ in the temperature range from 80 K to 400 K, finding an asymmertric line shape with a sharp high-energy cutoff and an exponentially decaying low-energy tail. In their work, the asymmetric line shape of WSe$_2$ at low temperatures was attributed to exciton localization instead of PSBs, because the authors believe a phonon-assisted line shape should be with an exponentially decaying high-energy tail and a sharp low-energy cutoff~\cite{Qi}, a spectral feature that is often found in absorption spectra.
However, the opposite is true in the PL spectra due to the mirror image relationship between the absorption and emission spectra~\cite{chr,bre}. Therefore, PSBs can not be excluded as a cause of the spectral asymmertry.
Moreover, as will be demonstrated in this work, our MBO model can effectively account for this line shape asymmetry in temperature-dependent emission spectra of WSe$_2$, by attributing it to PSBs. It is our belief that the MBO model is a powerful tool for investigating phonon effects on the linear optical spectra of TMDs and extracting essential parameters for further calculations of materials properties.

The remainder of the paper is structured as follows. In Sec.~\ref{MM}, the details of MBO model and the fitting procedure are given. In comparison with measurements,  the emission spectra of WSe$_2$,  WS$_2$, MoSe$_2$ and MoS$_2$  are fitted with the MBO model at various temperatures in Sec.~\ref{WSe2},~\ref{WS2},~\ref{MoSe2} and \ref{MoS2}, respectively. In Sec.~\ref{var}, the electron-phonon interaction strength in MBO model and semi-empirical Varshni equation is connected. Conclusions are drawn in Sec.~\ref{Con}.

\section{Model and Method}
\label{MM}
\subsection{The Brownian Oscillator Model}

In the MBO model~\cite{muk}, the system Hamiltonian is given by
\begin{align}\label{1}
H = |g\rangle H_g\langle g| + |e\rangle H_e\langle e| + H^{'}
\end{align}
where $|g\rangle$ and $|e\rangle$ represent the ground state and the excited state in a two-level system, respectively, and
\begin{align}\label{2}
H_g = \sum_{j}\left[\frac{p_j^2}{2m_j}+\frac{1}{2}m_j{\omega}_j^{2}q_j^{2}\right]
\end{align}
\begin{align}\label{3}
H_e = \hbar \omega_{eg}^0 + \sum_{j} \left[\frac{p_j^2}{2m_j}+\frac{1}{2}m_j{\omega}_j^{2}\left(q_j+d_j\right)^{2}\right]
\end{align}
and
\begin{align}\label{4}
H^{'} = \sum_{n} \left[\frac{P_n^2}{2m_n}+\frac{1}{2}m_n{\omega}_n^{2}\left(Q_n-\sum_j \frac{c_{nj}q_j}{m_n\omega_n^2}\right)^{2}\right]
\end{align}
In Eqs.~(\ref{1})-(\ref{4}), $p_j \left(P_n\right)$, $q_j \left(Q_n\right)$, $m_j \left(m_n\right)$ and $\omega_j \left(\omega_n\right)$ are the momentum, coordinate, mass and frequency of the $j$th  ($n$th) nuclear mode of the major (reservoir) oscillators, separately. $d_j$ is the dimensionless displacement of the $j$th mode in the electronic excited state. In Eq.~(\ref{3}), the $\hbar \omega_{eg}^0$ term is the ZPL energy in the TLS. $H^{'}$ represents the coupling between the primary modes and bath modes with a coupling strength $c_{nj}$. In Eq.~(\ref{4}), the damping effect is described by the cross term ${q_j}{Q_n}$. The energy gap operator is defined as
\begin{align}\label{5}
U = \sum_{j} U_j=H_e - H_g - \hbar \omega_{eg}^0
\end{align}

In addition to an expansion in eigenstates and semiclassical method of requiring the calculation of classical trajectories, the cumulant expansion is the third method to calculate the spectral response function, whose second order truncation can compute the linear spectra of absorbance and photoluminescence effectively. To obtain an response expression of linear spectra,  the correlation function of the $j$th primary mode can be defined as:
\begin{align}\label{6}
C_j \left(t \right) = -\frac{1}{2 \hbar^2}\left[\langle U_{j}(t)U_{j}(0)\rho_g \rangle- \langle U_{j}(0)U_{j}(t)\rho_g \rangle\right]
\end{align}
Here the term $U(t)$ is the operator of interaction representation, also known as the electronic energy gap in the Heisenberg picture with respect to ground state dynamics. $\langle U_{j}(t)U_{j}(0)\rho_g \rangle$ represents the expectation value for the operator of electronic energy gap of mode $j$ at $t$ and the operator at $t = 0$ acting on the ground-state vibrational density matrix. $C_j$ is the key quantity carrying all the necessary microscopic information for computing the optical response functions based on the second-order cumulant approximation. $\rho_g$ is the equilibrium ground-state vibrational density matrix written as
\begin{align}\label{7}
\rho_g = \frac{|g\rangle \langle g|  {\rm exp}(-\beta \hat{H_g} )}{Tr \left[{\rm exp}(-\beta \hat{H_g} )\right]}
\end{align}
where $\beta = \frac {1}{k_{\rm B}T}$. Eq.~(\ref{6}), the correlation function in time domain can be converted to frequency domain by the Fourier transform, then calculated by the path integral techniques, whose imaginary part is also known as the spectral density
\begin{eqnarray}\label{8}
\tilde C_j^{''} (\omega) = \frac {2\lambda _j \omega_j^2 \omega \gamma_j (\omega)}{\omega^2 \gamma_j^2 (\omega)+\left[\omega_j^2 + \omega \sum_j (\omega)-\omega^2 \right]^2}
\end{eqnarray}
Here $\gamma_j $ is the damping coefficient~\cite{dam} adjusting the smoothness of the curve, widths of the ZPL and the PSBs. $\sum_j (\omega)$ represents the real part of the self-energy~\cite{dam} and 2$\lambda_j$ is the contribution of the $j$th primary mode to the Stokes shift
\begin{align}\label{9}
2\lambda_j = \frac{m_j \omega_j^2 d_j^2}{\hbar}
\end{align}
Instead, Eq.~(\ref{9}) can be written as $\lambda_j = S_j \hbar \omega_j$, where $S_j$ is the dimensionless HR factor identifying the exciton-phonon interaction strength. In our work, a simple form of the MBO model is used, where the spectral distribution function $\gamma_j (\omega)$ is regarded as in its Ohmic limit (i.e., $\gamma_j (\omega)$ is a constant) and $\sum_j (\omega)$ is set as zero.   Thus the  spectral density function of the $j$th primary oscillators can be simplified to be the following form
\begin{eqnarray}\label{10}
\tilde C_j^{''} (\omega) = \frac {2\lambda _j \omega_j^2 \omega \gamma_j (\omega)}{\omega^2 \gamma_j^2 (\omega)+\left(\omega_j^2-\omega^2 \right)^2}
\end{eqnarray}

According to Eq.~(\ref{10}), the spectral response function $g(t)$ can be described as
\begin{eqnarray}\label{11}
g(t)=-\frac{1}{2\pi}\int_{-\infty}^{\infty} \,d\omega \frac{C^{''}(\omega)}{\omega^2}\left[1+{\rm coth}(\beta \hbar \omega/2)\right]&& \nonumber \\
( e^{-i\omega t}+i\omega t-1)&&
\end{eqnarray}

Then $C^{''}(\omega)$ and $\lambda$ can be expressed as the summation of individual contribution from each separate primary oscillator
\begin{align}
C^{''}(\omega) &= \sum_{j}C_j^{''}(\omega) \label{12} \\
\lambda &= \sum_{j} \lambda_j \label{13}
\end{align}
From Eqs.~(\ref{11})-(\ref{13}), the calculation of PL line shape follows
\begin{align}\label{14}
I_{\rm PL}(\omega) = \frac{1}{\pi}{\rm Re}\int_{0}^{\infty}{\rm exp}\left[ i(\omega - \omega_{eg} + \lambda)t-g^{*}(t)\right]\,dt
\end{align}

\subsection{The Fitting Procedure}
The frequencies of main phonon modes coupled to electronic excitations of interest are extracted from the Raman spectra of the specific TMDs monolayers, which ensures realistic TMD vibrational modes are considered regardless of materials preparation methods. One primary phonon mode is initially considered to couple to the secondary-phonon bath for {simplicity}. During the fitting of temperature-dependent PL, the primary phonon frequency is kept constant and the related HR factors are allowed to fluctuate. If one phonon mode is {insufficient}, then two primary modes combining a high-frequency mode $\omega_1$ and a low-energy mode $\omega_2$ (with a moderate $S_2$) are used, as the MBO fitting with one primary phonon mode usually leads to very narrow spectral peaks, rendering a low-energy phonon necessary to account for measured spectral line widths.
Traditional Brown-oscillator models often focus on one TLS. In our work, two or three independent TLSs are assumed depending on the details of emission spectra. A weighting factor to account for the contribution of each TLS as well as the corresponding transition dipole moment is used to arrive at the total emission spectrum. For two TLSs, the weighting factors are determined by the following procedure: first fix positions of two ZPL peaks from measured spectra, then use the MBO model to map out the two corresponding phonon side bands, before determining the weighting  factors to minimize the MSE and achieve the best fit of the measured spectra.

Spectral fitting with the semi-empirical Varshni equation is conducted by using the curve fitting tool of MATLAB\_R2021.

\section{Results and Discussion}
\label{RD}
\subsection{MBO fitting of WSe$_2$}
\label{WSe2}

\begin{figure}[tbp]
\centering
\includegraphics[scale=0.45,trim=0 0 0 0, clip]{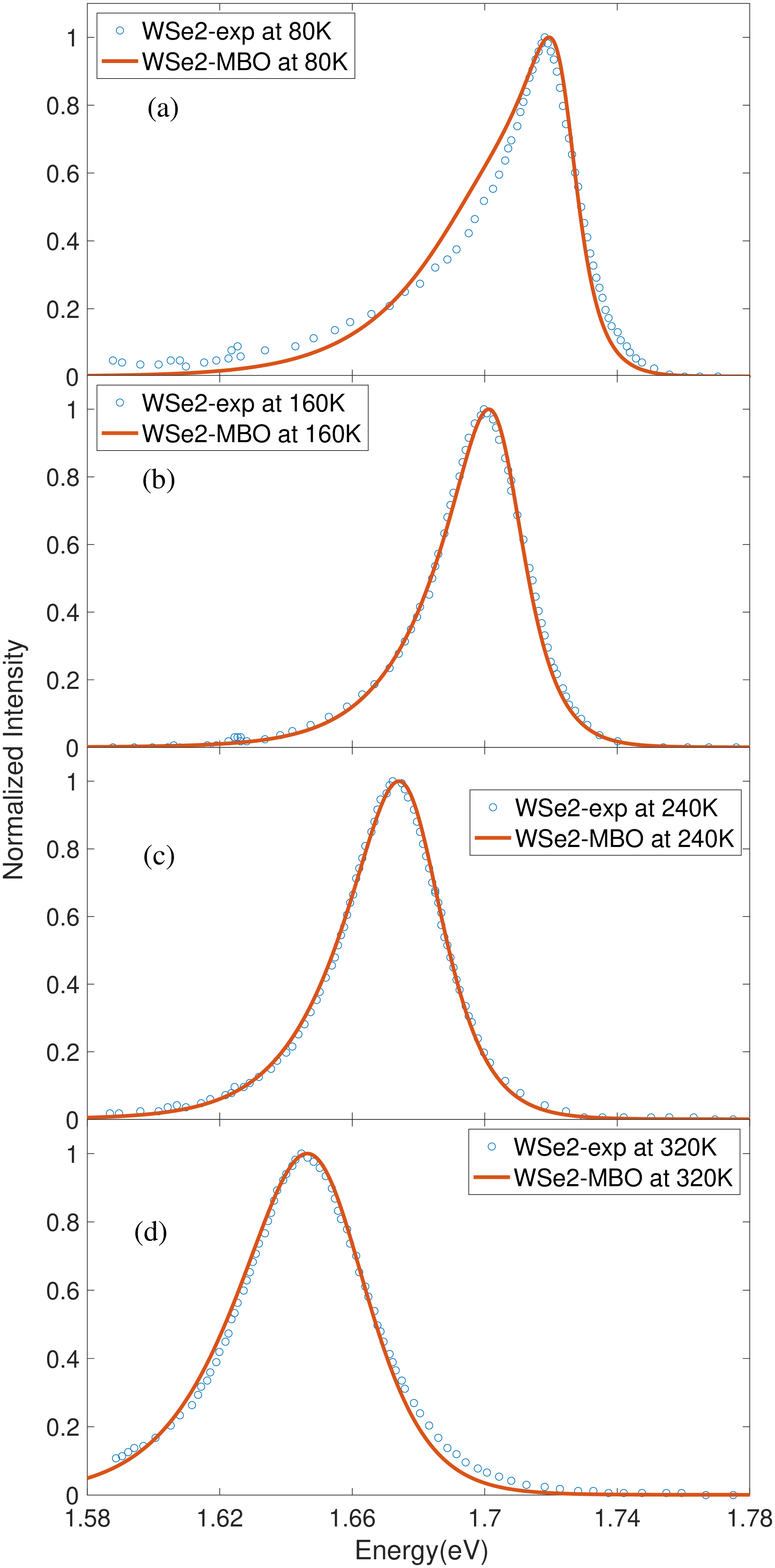}\\
\caption{Measured and MBO-fitted emission spectra of WSe$_2$ at different temperatures: (a) 80 K, (b) 160 K, (c) 240 K and (d) 320 K. MBO fitting parameters are collected in Table \ref{table1}.}
\label{fig1}
\end{figure}

{A typical TMD, WSe$_2$ has emission spectra with obvious phonon-assisted signatures resulting from an indirect band gap at lower temperatures such as 35 K~\cite{bre}. With the increasing temperature, more excitons could occupy the energetically higher bright state, which favors the direct emission process while reducing indirect PL signals. In Ref.~\cite{Qi}, the PL spectra of WSe$_2$ do not exhibit phonon-assisted signatures due to a higher range of temperatures, i.e., $80~\rm K\sim320~\rm K$, where there is only one broad peak which arises from the direct band gap.}
Shown in Fig.~\ref{fig1} are measured emission spectra of mechanically exfoliated WSe$_2$ at various temperatures~\cite{Qi} and their MBO simulation. The primary phonon frequency of about 250 cm$^{-1}$ (0.031 eV) is extracted from the Raman spectra~\cite{Qi}. For a given temperature, there is only one spectral peak which is increasingly more redshifted and symmetrical as the temperature increases, accompanied with spectral broadening.

For semiconductors, as the temperature increases, the crystal lattice dilates and the interatomic bonds are weakened, and less energy is required for electrons to hop to conduction bands, resulting in a decreased band gap. TMD monolayers are atomically thin 2D-semiconductors, which respond similarly to temperature variations. The MBO model can accurately describe the sidebands of two primary modes (with frequencies 250 and 100 cm$^{-1}$) on the WSe$_2$ emission spectrum, together with the dissipative effect of the secondary phonon bath. The high-frequency mode introduces broadening that is related to phonon side bands (PSBs), controlling the overall  spectral asymmetry, which is visible in Fig. 1. The effect of the low-frequency mode~\cite{selig} lies in the broadening of ZPL peak, which is less visible in the PL spectrum because of its small Huang-Rhys factor.
Obtained fitting parameters of WSe$_2$ are given in Table \ref{table1}.
The corresponding HR factors of two primary oscillators are 0.70 and 0.10 at 80 K in Fig. \ref{fig1}$(a)$.
The small deviation between the measured PL spectrum and its MBO fitting at 80K can be attributed to the broaden effects from the weak contribution of the indirect band gap~\cite{bre}.

As the temperature rises from 160 K to 320 K, the spectral peaks are red-shifted and become more symmetrized, as shown in Figs.~\ref{fig1}(\rm{b}), (\rm{c}) and (\rm{d}). Despite the temperature variation, the HR factors of the two primary oscillators are found to be 0.33 and 0.05. {Sudden decreases in the HR factors, similar to the drop from (0.70, 0.10) for excitons at 80 K to (0.33, 0.05) for excitons at 320 K}, have been previously reported for the polar semiconductor ZnSe \cite{zhao}.

\begin{table}[tbp]
\centering
\topcaption{Fitted MBO parameters for the PL spectra of WSe$_2$ at different temperatures. The units of $\omega$ and $\gamma$ are cm$^{-1}$, and that of $\hbar\omega_{eg}$ and T are eV and K, separately.}
\label{table1}
\begin{tabular}{| c | c | c | c | c | c | c | c |}
\hline
$\omega_1$ & $\omega_2$ & $\gamma_1$ & $\gamma_2$ & $S_1$ & $S_2$ & $\hbar\omega_{eg}$ & T \\\hline
250 & 100 & 282 & 282 & 0.70 & 0.10 & 1.724 & 80 \\\hline
250 & 100 & 403 & 403 & 0.35 & 0.05 & 1.705 & 160 \\\hline
250 & 100 & 403 & 403 & 0.30 & 0.05 & 1.678 & 240 \\\hline
250 & 100 & 403 & 403 & 0.33 & 0.05 & 1.652 & 320 \\\hline
\end{tabular}
\end{table}

\subsection{MBO fitting of WS$_2$}
\label{WS2}
\begin{figure}[tbp]
\centering
\includegraphics[scale=0.5,trim=0 0 0 0, clip]{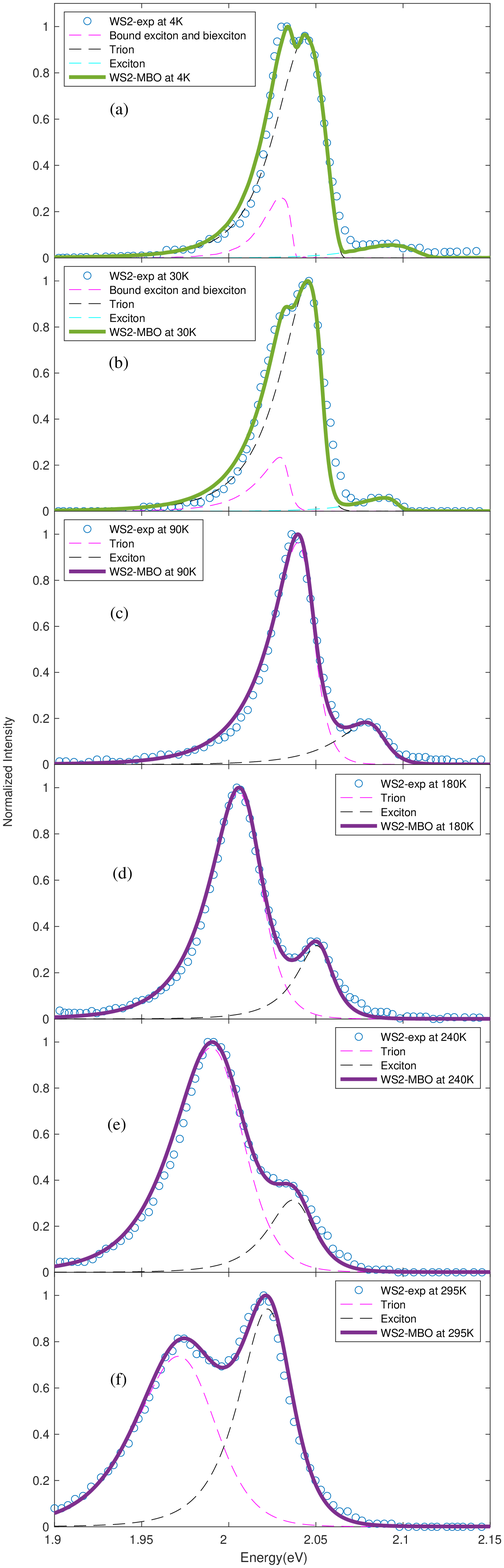}\\
\caption{Measured and MBO-fitted emission spectra of WS$_2$ at different temperatures: (a) 4 K, (b) 30 K, (c) 90 K, (d) 180 K, (e) 240 K and (f) 295 K. MBO fitting parameters are collected in Table \ref{table2}.}
\label{fig2}
\end{figure}

WS$_2$ monolayers are known to exhibit more complex emission spectra~\cite{miti, zhu, pei, bel}, which involve biexciton states at low temperatures and a stable trion state existing in the temperature range {from 4 K to 295 K~\cite{ref17}}. Displayed in Fig.~\ref{fig2} are measured~\cite{ref17} and simulated PL spectra  of mechanically exfoliated monolayer WS$_2$. As shown in Figs.~\ref{fig2} (\rm{a}) and (\rm{b}), at the low temperatures of 4 K and 30 K, there are three emission peaks: the lowest-energy peak is assigned to the superposition state of the defect-bound exciton ($\rm L_1$) and biexciton ($\rm XX$) emission, the adjacent higher-energy peak around 2.05 eV is attributed to the trion emission, and the highest-energy peak near 2.1 eV is due to the exciton emission \cite{ref17}. Despite the contention still surrounding the lowest-energy spectral shoulder, the measured spectra can be fully accounted for by the MBO model.

The spectra at temperatures from 90 K to 295 K are shown in Figs.~\ref{fig2} (\rm{c})-(\rm{f}). There are two major peaks in the emission spectra. The low-energy peak is identified as the trion emission~\cite{ref17, zhu, pei}, and the other one is recognized as the exciton state, gaining intensities as well as redshifts as the temperature increases. The trion states have much higher emission intensities than those of the excitons until the temperature is increased to 295 K where the two switch places. For simplicity, two (three) independent TLSs are considered in the MBO model for spectral fitting at high (low) temperatures.

Excellent agreement is found between the MBO simulation and measured PL spectra~\cite{ref17}, which confirms the validity of the MBO model here. Fitting parameters are listed in Table \ref{table2}. The primary phonon frequency about 352 cm$^{-1}$ (0.044 eV) is extracted from the Raman spectra~\cite{ref18} and only one primary mode is used for the relaxation of trion, exciton, $\rm L_1/XX$. For a given temperature the same $\gamma$ is assumed.  A MBO model that includes three TLSs is used for the WS$_2$ at 4 K and 30 K to incorporate the $\rm L_1/XX$ superposition state.
$\hbar\omega_{eg1}$, $\hbar\omega_{eg2}$ and $\hbar\omega_{eg3}$ are the ZPL energies for the $\rm L_1/XX$ superposition state, the trion state, and the exciton state, respectively. The corresponding HR factors are $S_1$, $S_2$ and $S_3$. At 30K, $S_1$ is constant at 0.4 different from the dramatic decrease of $S_2$ and $S_3$.
When temperature falls in the range between 90 K and 295 K, {$S_1$ disappears because increased thermal energy frees trapped excitons~and breaks up biexcitons \cite{ref17, zhu}.} It is found that $S_2$ and $S_3$ oscillate around 0.35 and 0.2, respectively. $\alpha_1$, $\alpha_2$ and $\alpha_3$ label the individual emission contributions from the $\rm L_1/XX$,  the trion and the exciton state. The share of total PL line shape is shifted from trion to exciton part gradually. Moreover, in the Table \ref{table2} the term $\hbar\omega_{eg3}-\hbar\omega_{eg2}$ fluctuate little around 45 meV, suggesting the trion binding energy is hardly affected by temperature. This trion binding energy is reasonable since exciton binding energy is usually one order larger than trion's and the reported exciton binding energy~\cite{ref19, ref20, ref21, ref22} for single-layer WS$_2$ is in the order of 0.5 eV.

\begin{table*}[tbp]
\centering
\topcaption{Fitted MBO parameters for the PL spectra of WS$_2$ at various temperatures. The units of $\omega$ and $\gamma$ are cm$^{-1}$, and that of $\hbar\omega_{eg}$ and T are eV and K, separately.}
\label{table2}
\label{table:2}
\begin{tabular}{| c | c | c | c | c | c | c | c | c | c | c | c | c |}
\hline
$\omega$ & $\gamma$ & $S_1$ & $S_2$ & $S_3$ & $\hbar\omega_{eg1}$ & $\hbar\omega_{eg2}$ & $\hbar\omega_{eg3}$ & $\hbar\omega_{eg3}-\hbar\omega_{eg2}$ & $\alpha_1(\%)$ & $\alpha_2(\%)$ & $\alpha_3(\%)$ & T \\\hline
352 & 2420 & 0.40 & 0.80 & 1.00 & 2.038 & 2.067 & 2.124 & 0.057 & 13.12 & 81.42 &  5.46 & 4 \\\hline
352 & 1210 & 0.40 & 0.53 & 0.50 & 2.034 & 2.054 & 2.097 & 0.043 & 14.67 & 80.89 & 4.44 & 30  \\\hline
352 & 807 & $\backslash$ & 0.40 & 0.45 & $\backslash$ & 2.045 & 2.086 & 0.041 & 0 & 83.05 & 16.95 & 90 \\\hline
352 & 807 & $\backslash$ & 0.30 & 0.20 & $\backslash$ & 2.011 & 2.053 & 0.042 & 0 & 80.77 & 19.23 & 180 \\\hline
352 & 807 & $\backslash$ & 0.35 & 0.20 & $\backslash$ & 1.998 & 2.040 & 0.042 & 0 & 82.14 & 17.86 & 240 \\\hline
352 & 807 & $\backslash$ & 0.35 & 0.20 & $\backslash$ & 1.980 & 2.026 & 0.046 & 0 & 52.38 & 47.62 & 295 \\\hline
\end{tabular}
\end{table*}

\subsection{MBO fitting of MoSe$_2$}
\label{MoSe2}

\begin{figure*}[tbp]
\centering
\subfigure[]{
\includegraphics[scale=0.16,trim=145 0 40 0]{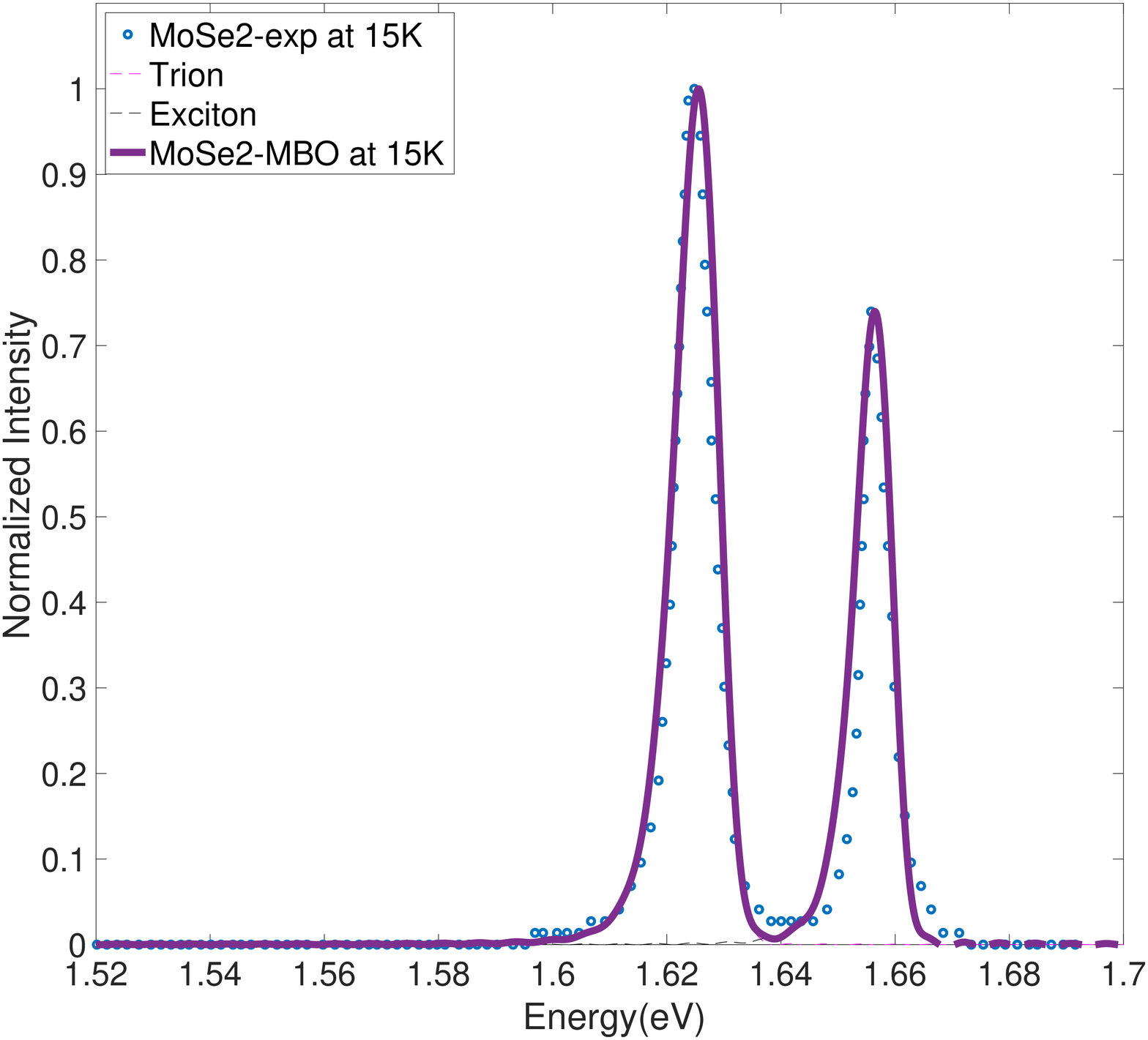}
}
\quad
\subfigure[]{
\includegraphics[scale=0.16,trim=100 0 40 0]{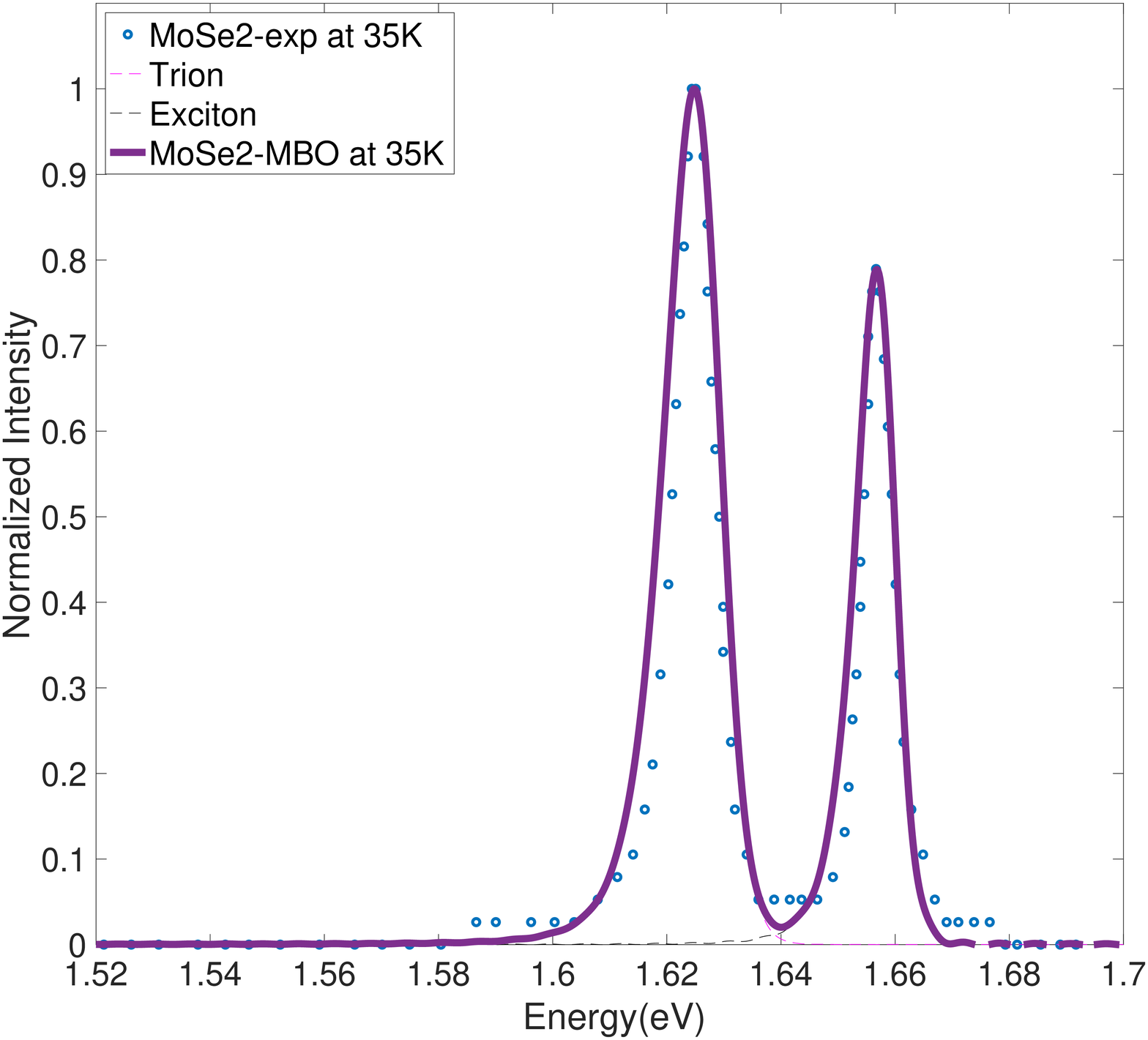}
}
\quad
\subfigure[]{
\includegraphics[scale=0.16,trim=100 0 40 0]{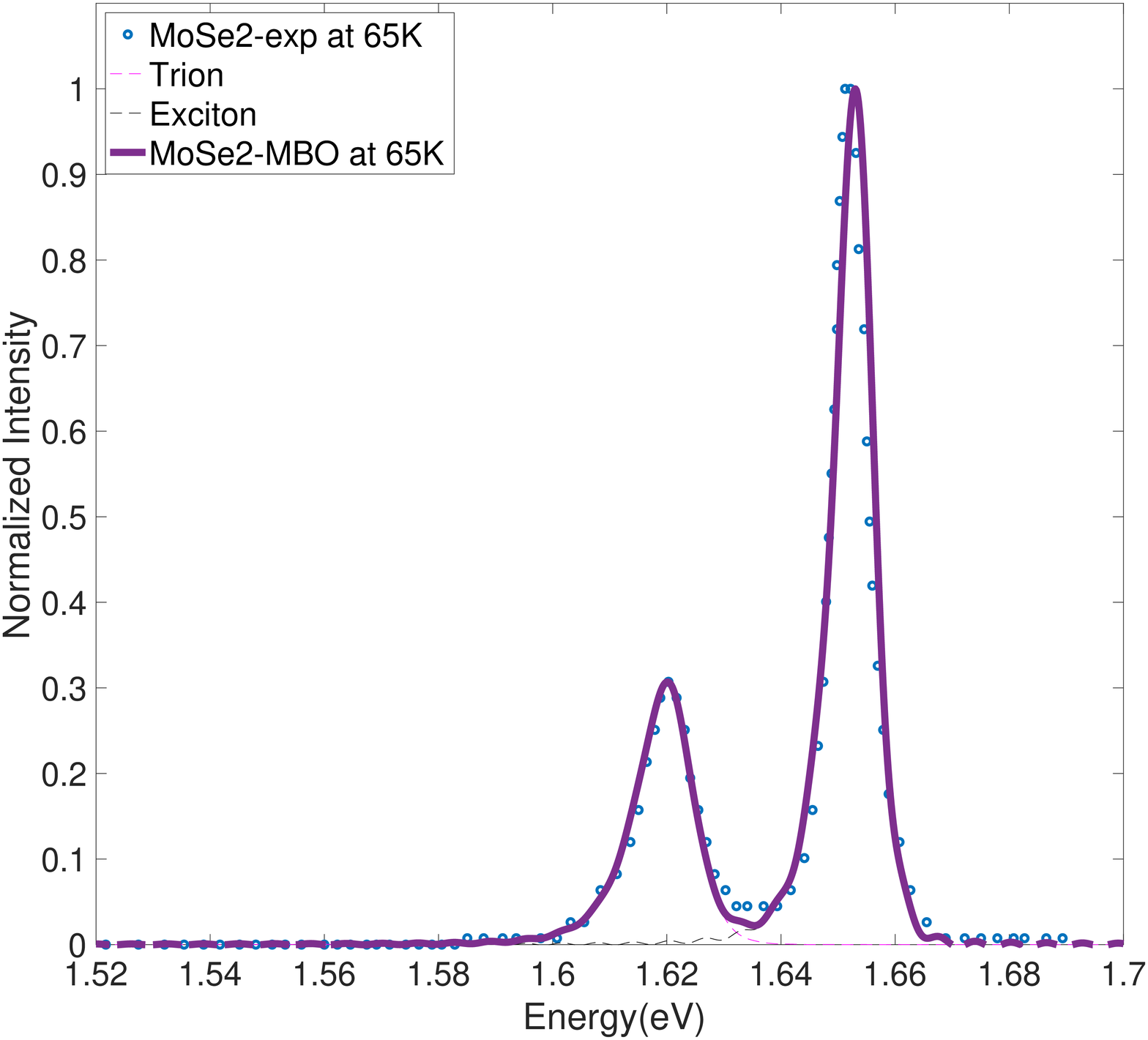}
}\\
\subfigure[]{
\includegraphics[scale=0.16,trim=145 0 40 0]{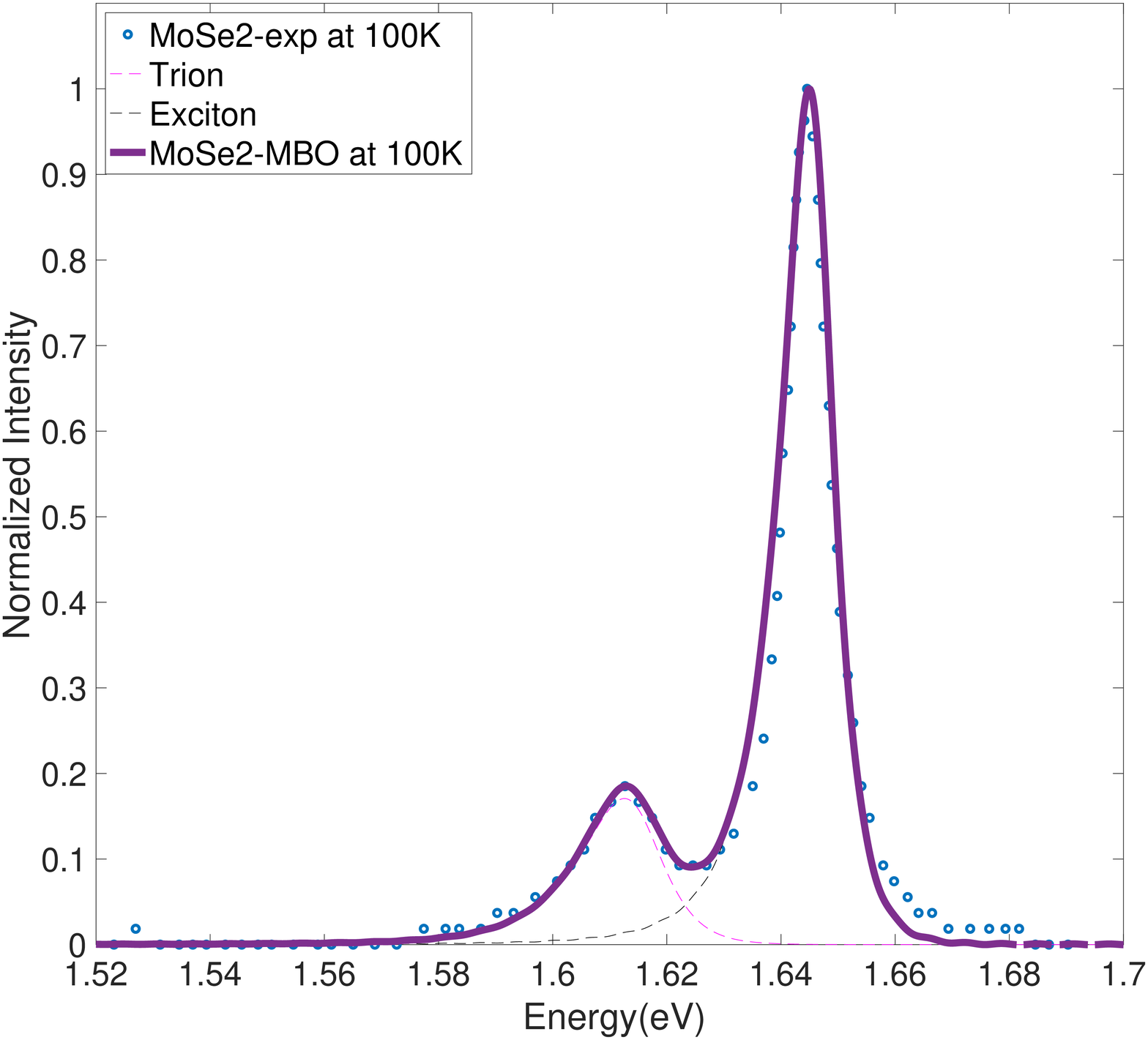}
}
\quad
\subfigure[]{
\includegraphics[scale=0.16,trim=100 0 40 0]{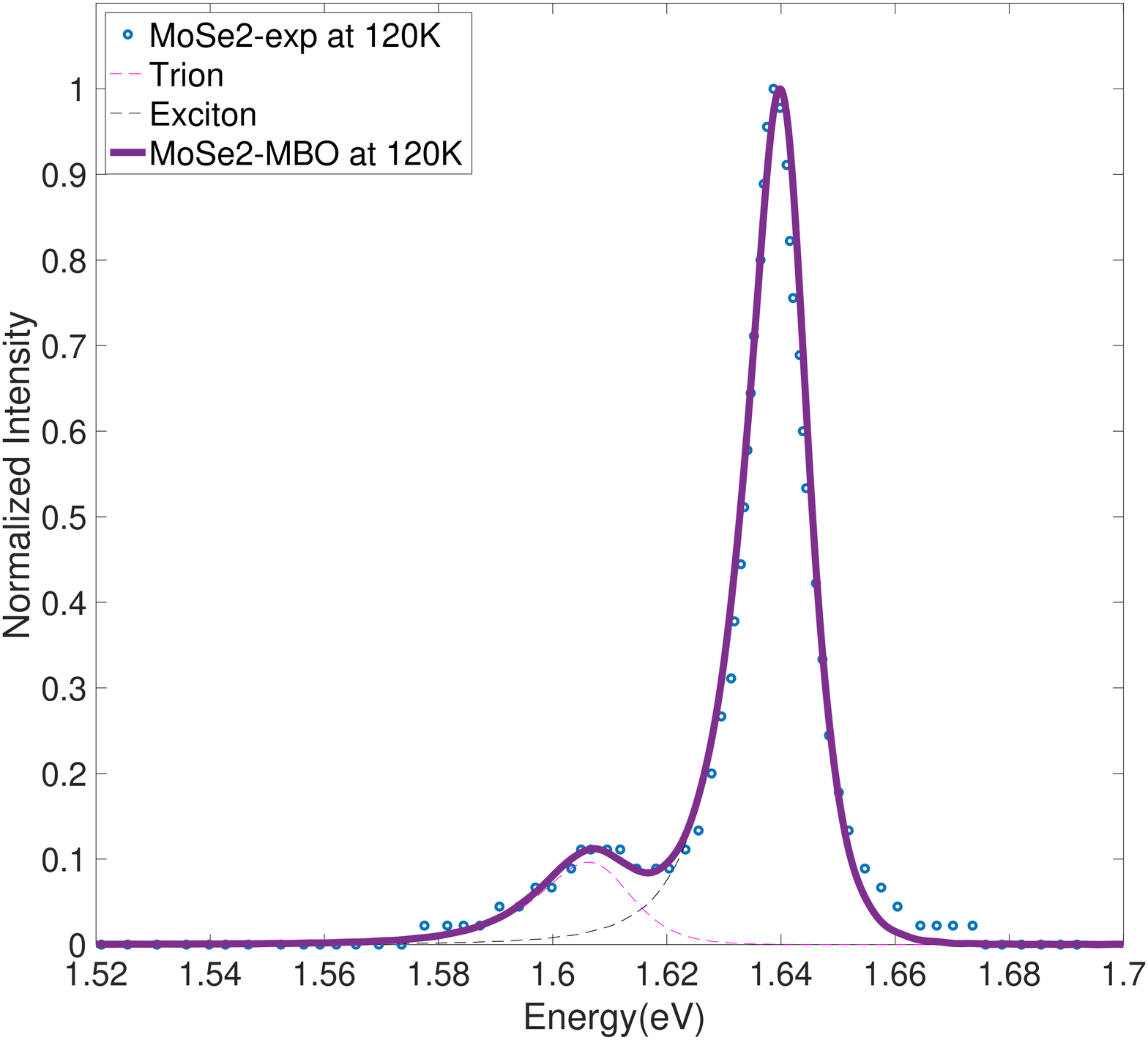}
}
\quad
\subfigure[]{
\includegraphics[scale=0.16,trim=100 0 40 0]{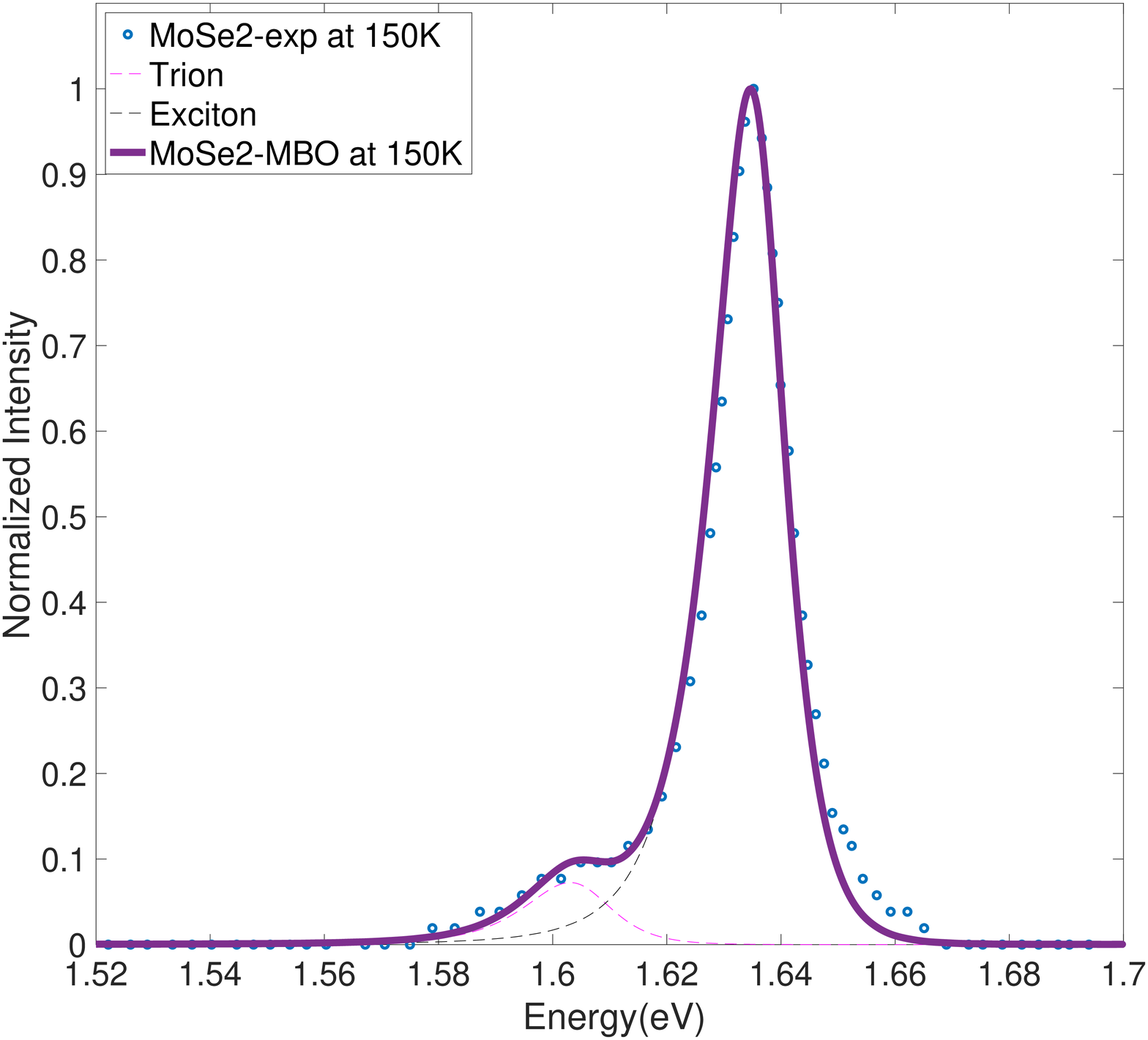}
}\\
\subfigure[]{
\includegraphics[scale=0.16,trim=145 0 40 0]{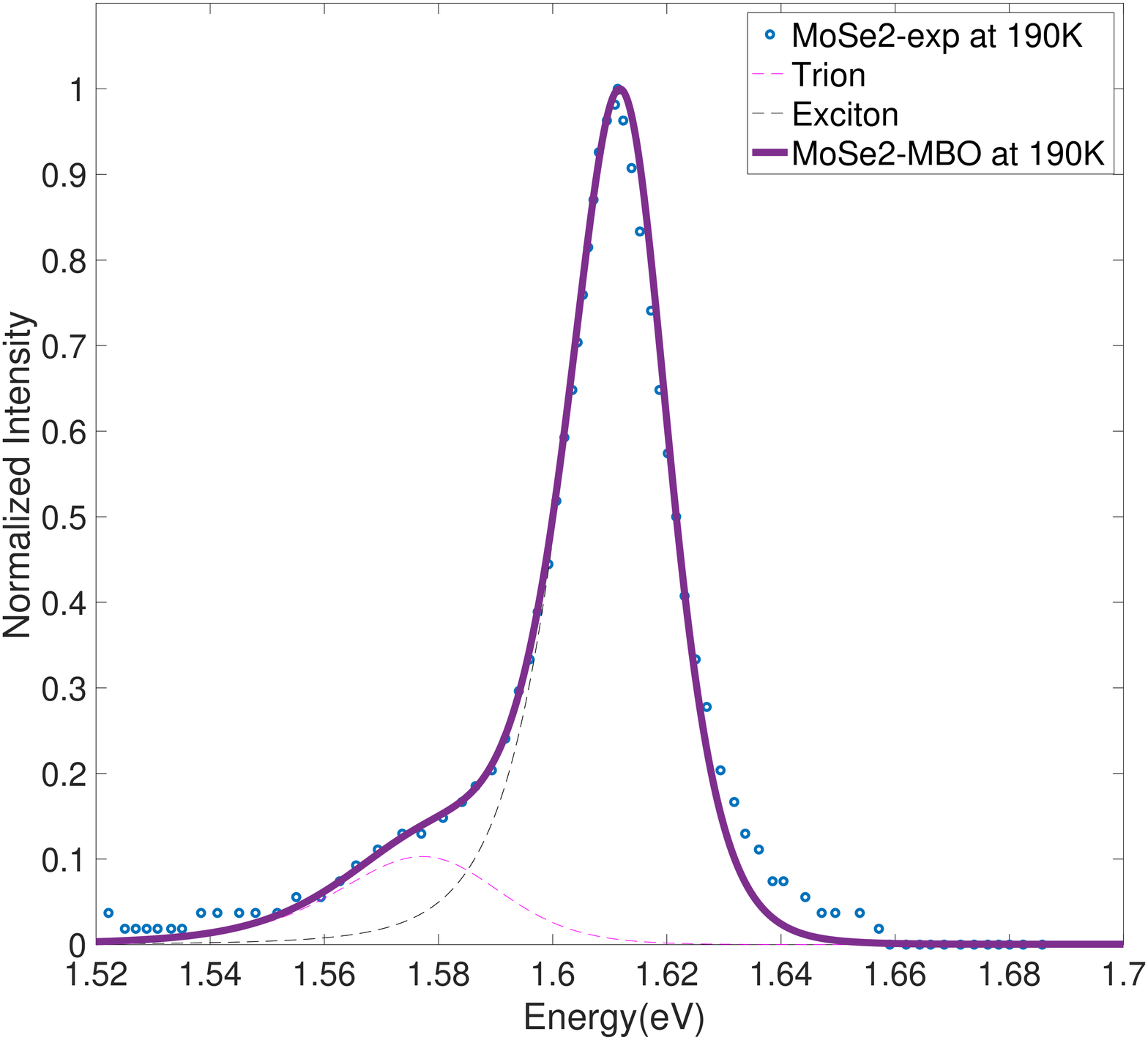}
}
\quad
\subfigure[]{
\includegraphics[scale=0.16,trim=100 0 40 0]{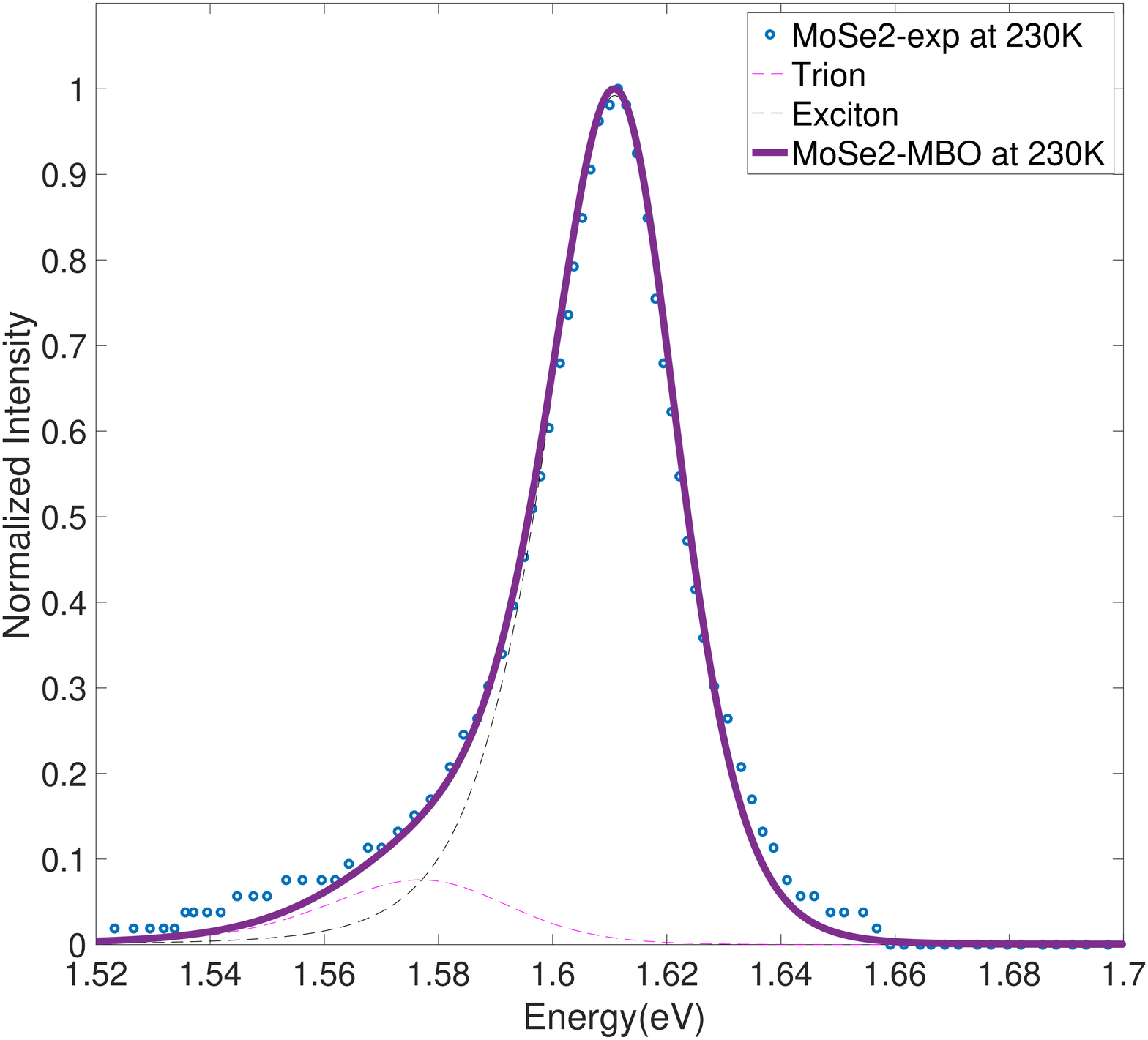}
}
\quad
\subfigure[]{
\includegraphics[scale=0.16,trim=100 0 40 0]{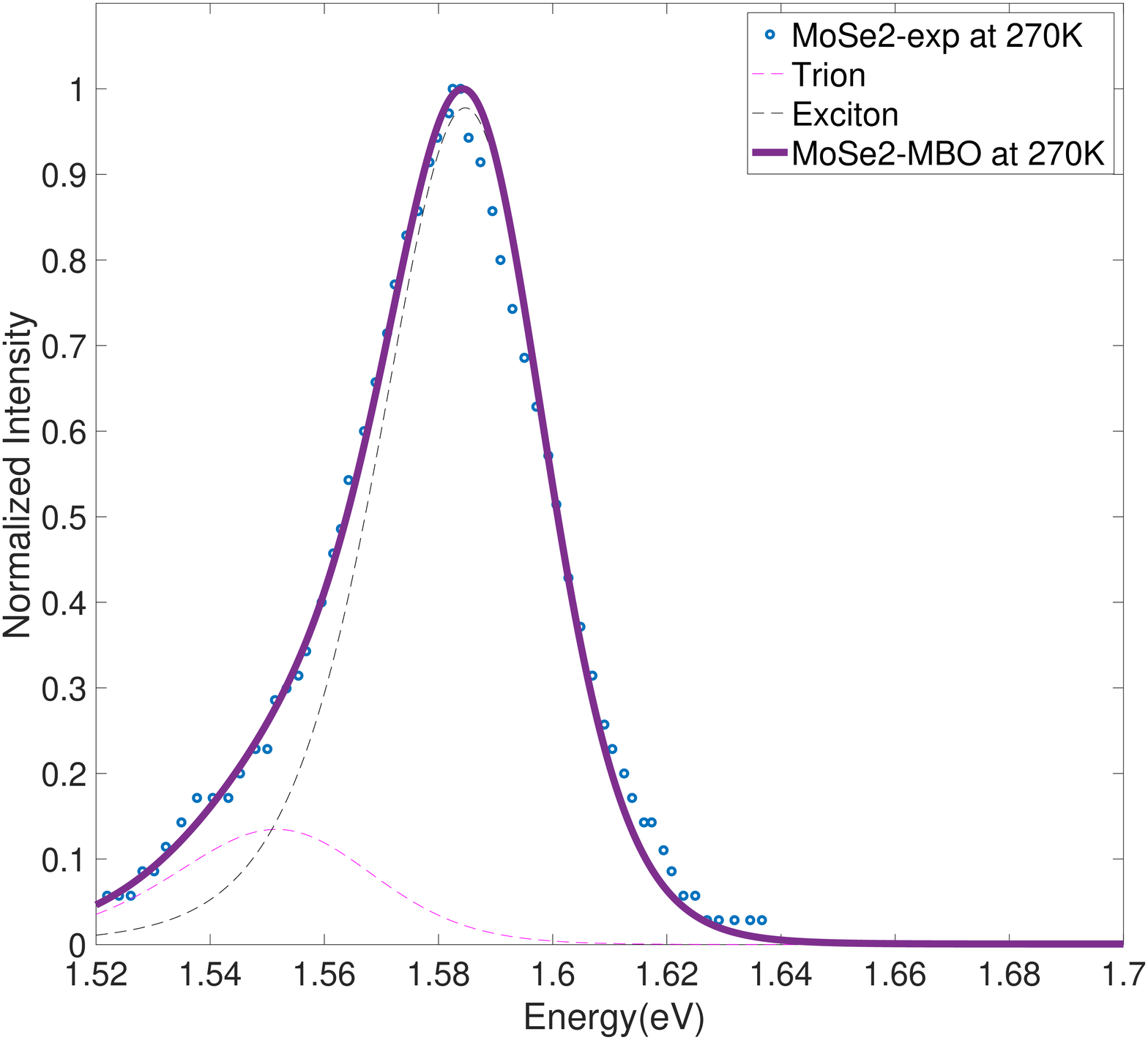}
}\\
\caption{Measured and MBO-fitted emission spectra of MoSe$_2$ at different temperatures: (a) 15 K, (b) 35 K, (c) 65 K, (d) 100 K, (e) 120 K, (f) 150 K, (g) 190 K, (h) 230 K and (i) 270 K. MBO fitting parameters are collected in Table \ref{table3}.}
\label{fig3}
\end{figure*}

Another 2D TMD semiconductor is the high-quality monolayer molybdenum diselenide (MoSe$_2$). Their temperature-dependent PL~\cite{ros} and the corresponding MBO fitting  are presented in Fig.~\ref{fig3}. The phonon mode of around 240 cm$^{-1}$ (0.03 eV) is extrated from the Raman spectra~\cite{ref23, ref24}. Similar to WS$_2$, in Fig.~\ref{fig3} the lower energy peak represents trion emission~\cite{mak1} and the higher one is exciton peak, which has been attributed to the valence band splitting induced by the spin-orbit coupling~\cite{evan}. At low temperatures such as 15 K and 35 K in Fig.~\ref{fig3}(\rm{a}) and (\rm{b}), trion emission occupies the majority of PL intensity and exceeds the exciton states. However, at 65 K in Fig.~\ref{fig3}(\rm{c}) PL intensity of trion state drops dramatically and its weight shifts to exciton emission gradually as temperature rises from 65 K to 270 K shown in Fig.~\ref{fig3}(\rm{d})-(\rm{i}).

Fitting parameters are listed in Table \ref{table3} and a two-TLS MBO model is employed with one common phonon mode frequency for the trion and exciton emission. Here $\alpha_1$ and $\alpha_2$ stand for the contribution percentage of the trion and exciton emission to the total line shape. $\hbar\omega_{eg1}$ and $\hbar\omega_{eg2}$ are the ZPL energies for the trion and exciton states, respectively. At 270 K, $\alpha_2$ around 83.47\% in Table \ref{table3} demonstrates PL of MoSe$_2$ under this condition is nearly only made up of exciton emission, which is distinguished from PL of WS$_2$ at 295K controlled by trion emission obviously with the $\alpha_2$ larger than 50\% in Table \ref{table2}. This implies the role of trion emission playing in the formation of PL line shape {varies in} TMDs. In the temperature range between 15 K and 270 K, the HR factors of the trion and exciton states oscillate around 0.22 ($S_1$) and 0.14 ($S_2$), respectively, insensitive to the temperature variation. {The trion binding energy, i.e., $\hbar\omega_{eg2}-\hbar\omega_{eg1}$, is kept constant around 31 meV, barely affected by the temperatures because the ZPL energies of the trion and the exciton have the nearly same variation in Table \ref{table3}.} Several PL spectra of MoSe2 reveal high-energy bands, which can be related to hot transitions, the Hezrberg-Teller effect, and the Duschinsky effect. Some of those effects can be incorporated by using more sophisticated methodologies~\cite{bor1,bor2,bor3,capo} for simulating spectral line shapes.

\begin{table*}[tbp]
\centering
\topcaption{Fitted MBO parameters for the PL spectra of MoSe$_2$ at various temperatures. The units of $\omega$ and $\gamma$ are cm$^{-1}$, and that of $\hbar\omega_{eg}$ and T are eV and K, separately.}
\label{table3}
\begin{tabular}{| c | c | c | c | c | c | c | c | c | c | c |}
\hline
$\omega$ & $\gamma$ & $S_1$ & $S_2$ & $\hbar\omega_{eg1}$ & $\hbar\omega_{eg2}$ & $\hbar\omega_{eg2}-\hbar\omega_{eg1}$ & $\alpha_1(\%)$ & $\alpha_2(\%)$ & T \\\hline
240 & 8066 & 0.30 & 0.23 & 1.633 & 1.662 & 0.029 & 61.27 & 38.73 & 15 \\\hline
240 & 3226 & 0.20 & 0.12 & 1.629 & 1.659 & 0.030 & 63.46 & 36.54 & 35 \\\hline
240 & 1613 & 0.13 & 0.08 & 1.622 & 1.654 & 0.032 & 28.57 & 71.43 & 65 \\\hline
240 & 807 & 0.20 & 0.12 & 1.615 & 1.646 & 0.031 & 15.97 & 84.03 & 100 \\\hline
240 & 807 & 0.20 & 0.12 & 1.609 & 1.641 & 0.032 & 9.91 & 90.09 & 120 \\\hline
240 & 807 & 0.15 & 0.12 & 1.605 & 1.636 & 0.031 & 9.09 & 90.91 & 150 \\\hline
240 & 807 & 0.30 & 0.15 & 1.583 & 1.614 & 0.031 & 12.28 & 87.72 & 190 \\\hline
240 & 807 & 0.30 & 0.18 & 1.583 & 1.614 & 0.031 & 13.04 & 86.96 & 230 \\\hline
240 & 807 & 0.30 & 0.22 & 1.558 & 1.589 & 0.031 & 16.53 & 83.47 & 270 \\\hline
\end{tabular}
\end{table*}

\subsection{MBO fitting of MoS$_2$}
\label{MoS2}
\begin{figure*}[tbp]
\centering
\subfigure[]{
\includegraphics[scale=0.16,trim=145 0 40 0]{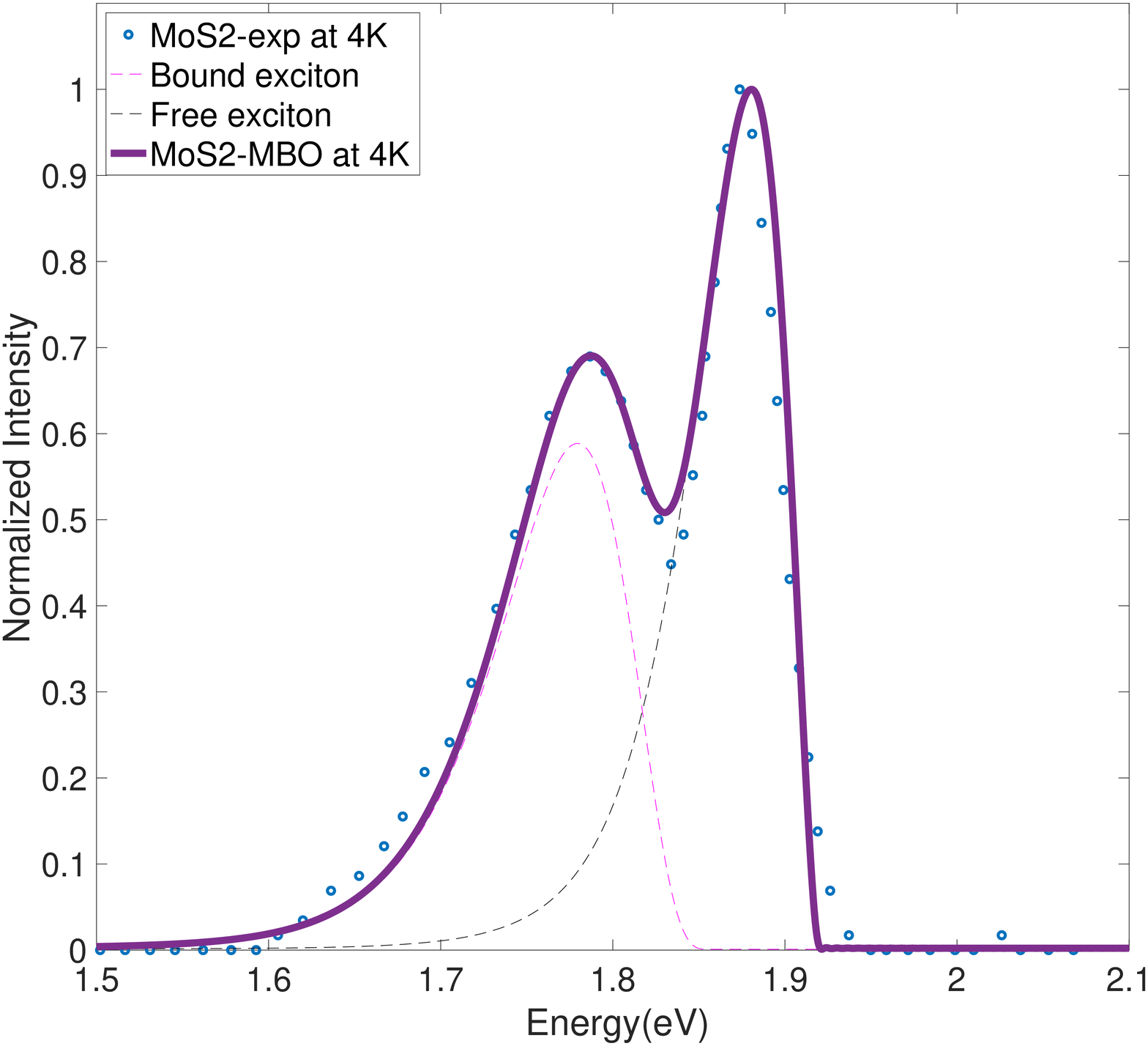}
}
\quad
\subfigure[]{
\includegraphics[scale=0.16,trim=100 0 40 0]{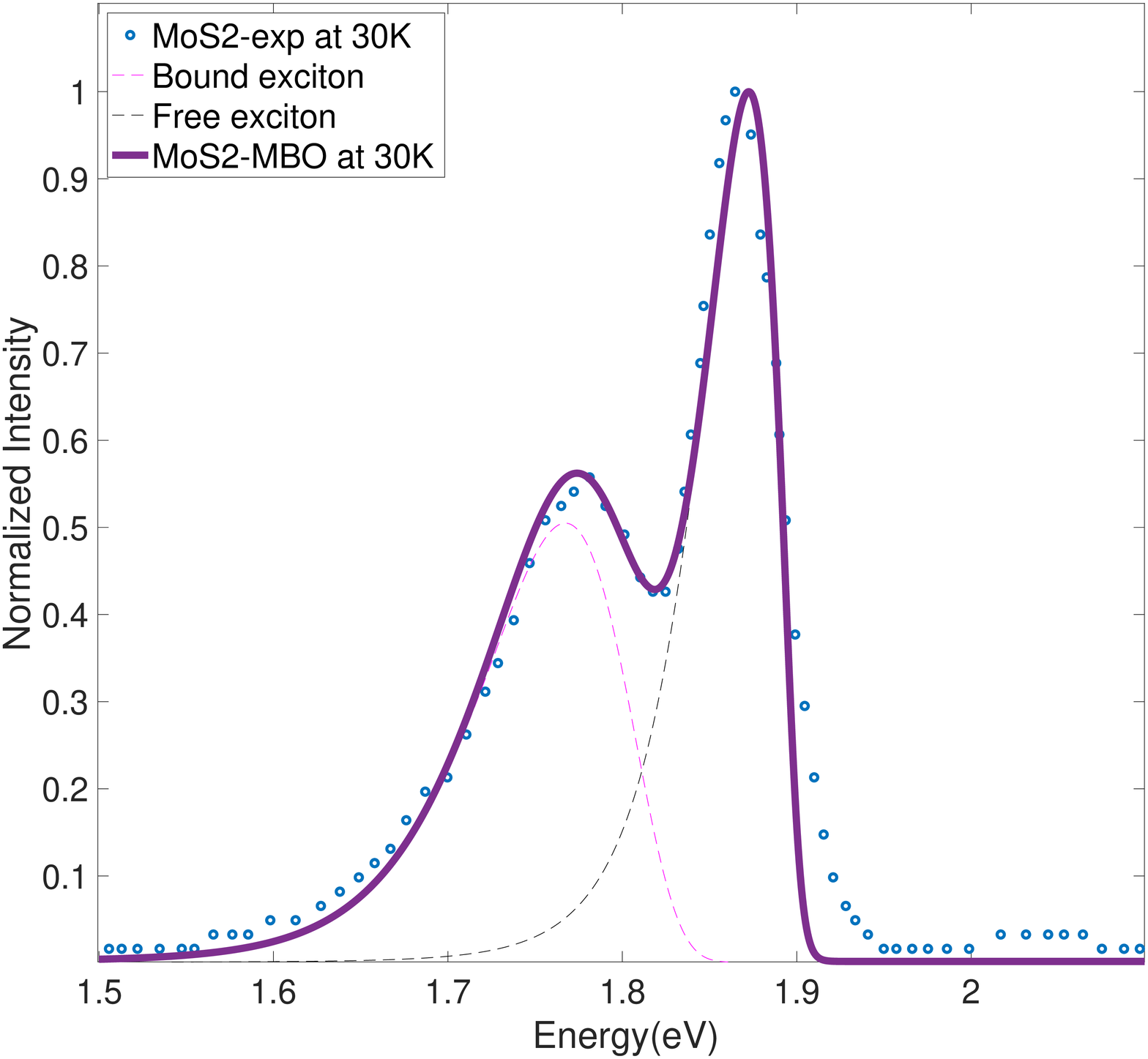}
}
\quad
\subfigure[]{
\includegraphics[scale=0.16,trim=100 0 40 0]{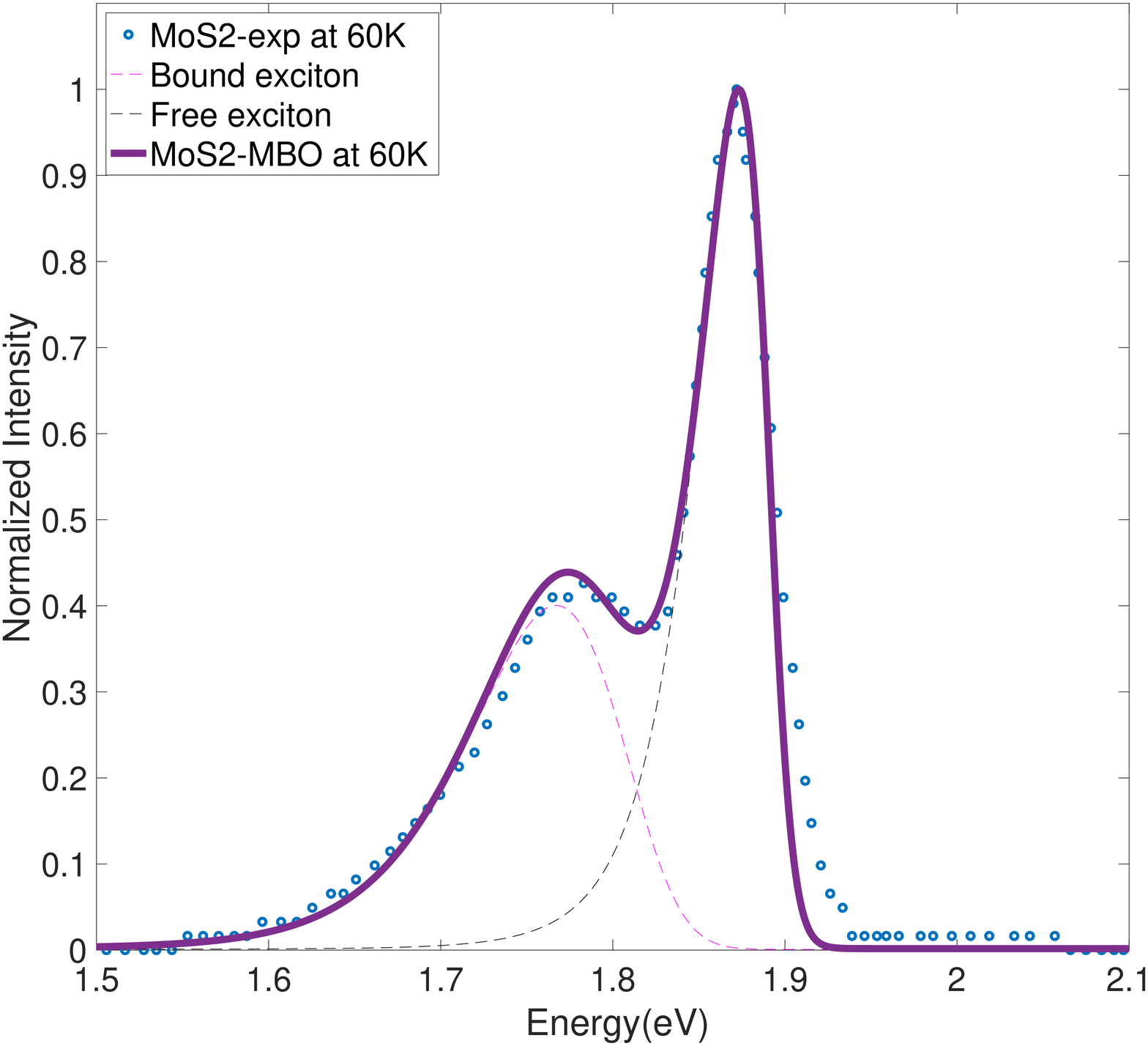}
}\\
\subfigure[]{
\includegraphics[scale=0.16,trim=145 0 40 0]{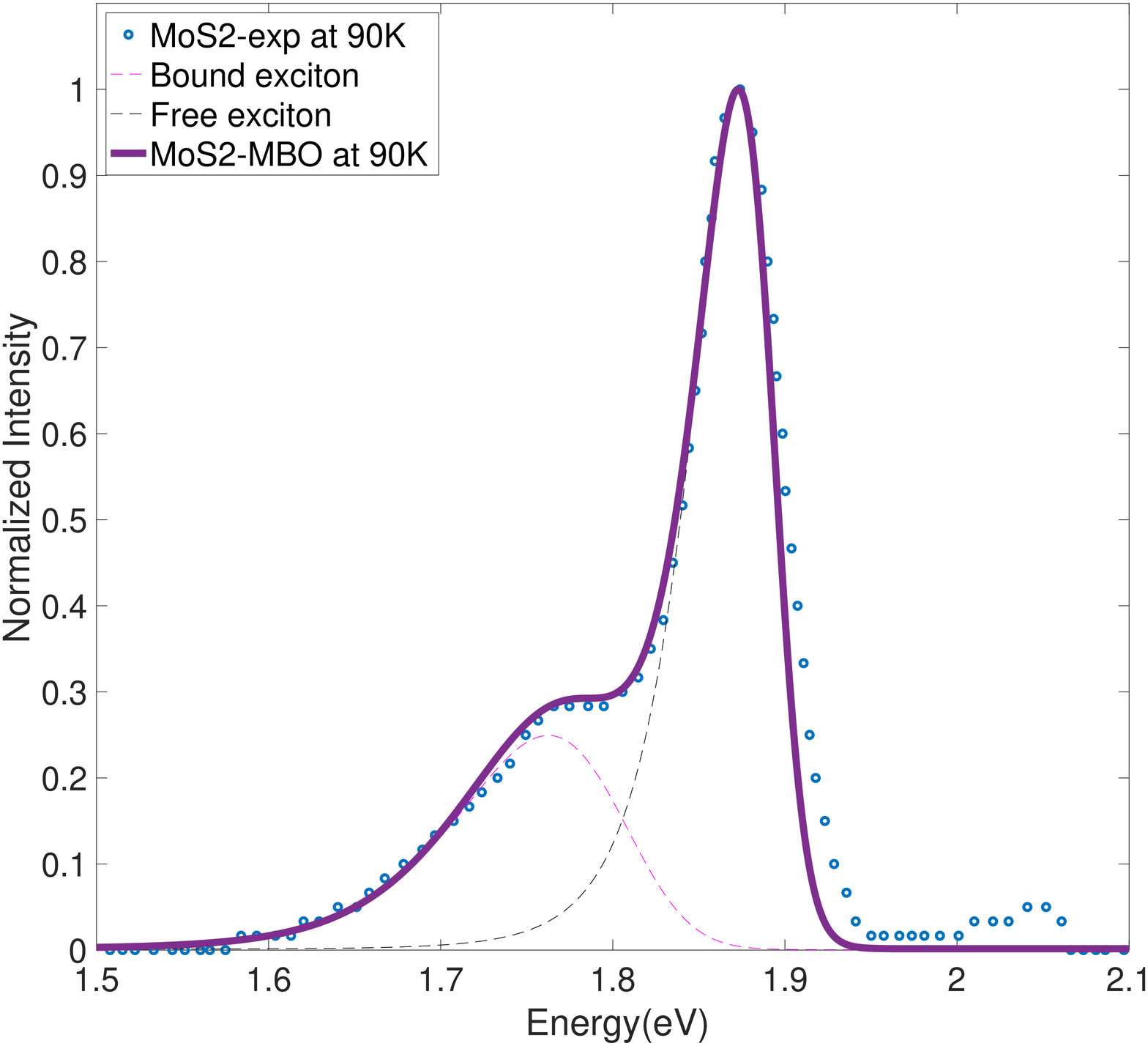}
}
\quad
\subfigure[]{
\includegraphics[scale=0.16,trim=100 0 40 0]{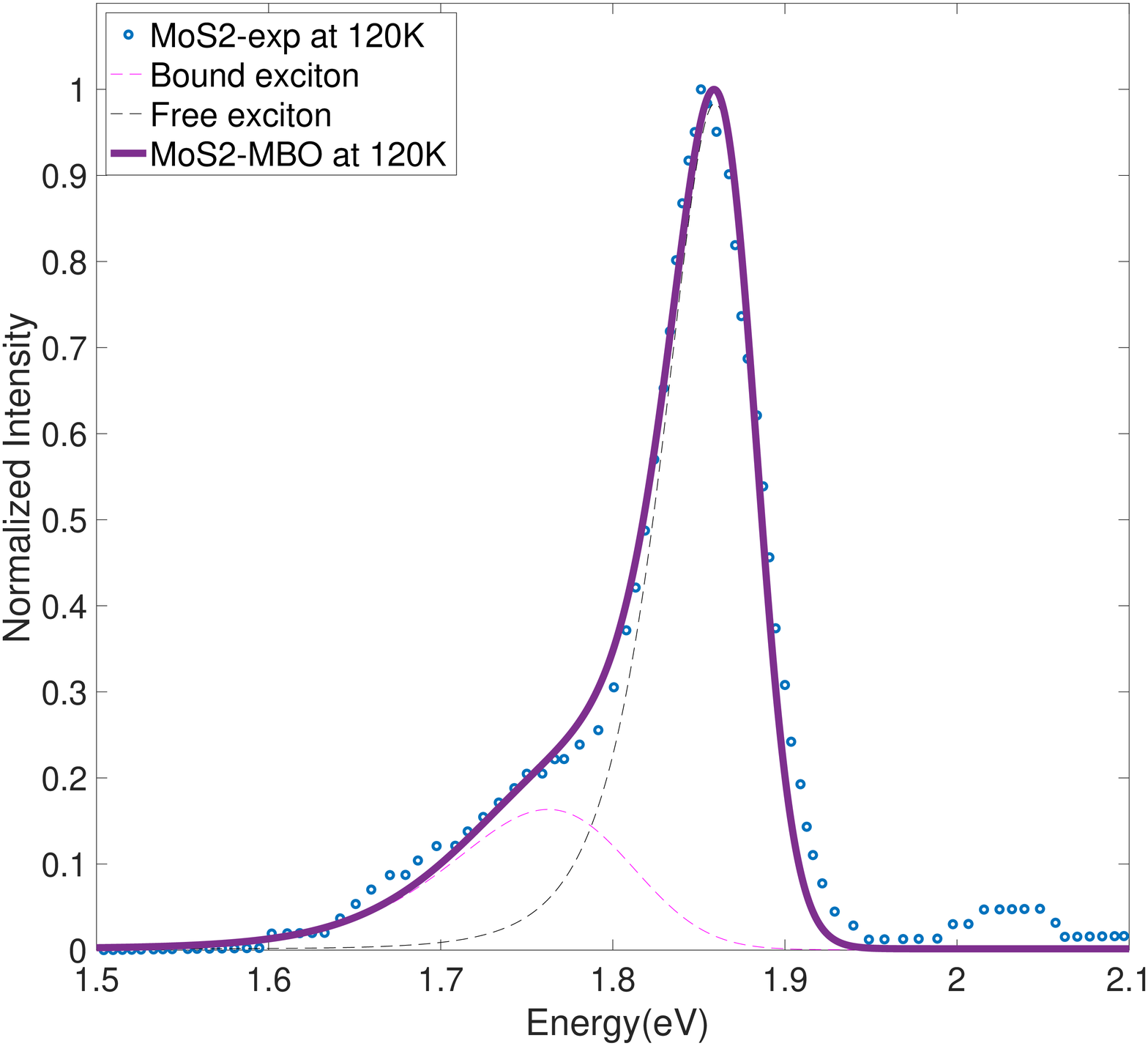}
}
\quad
\subfigure[]{
\includegraphics[scale=0.16,trim=100 0 40 0]{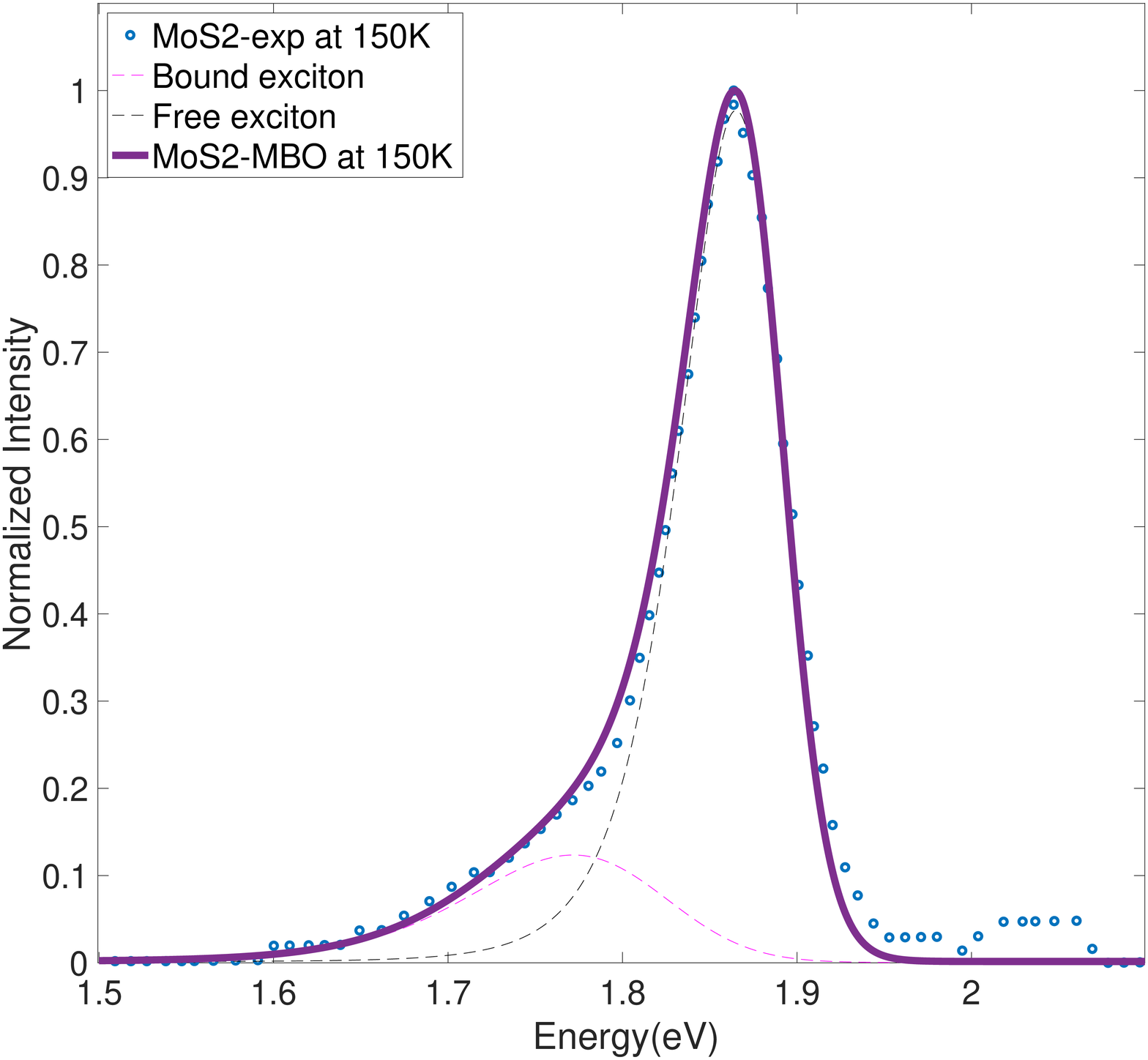}
}\\
\subfigure[]{
\includegraphics[scale=0.16,trim=145 0 40 0]{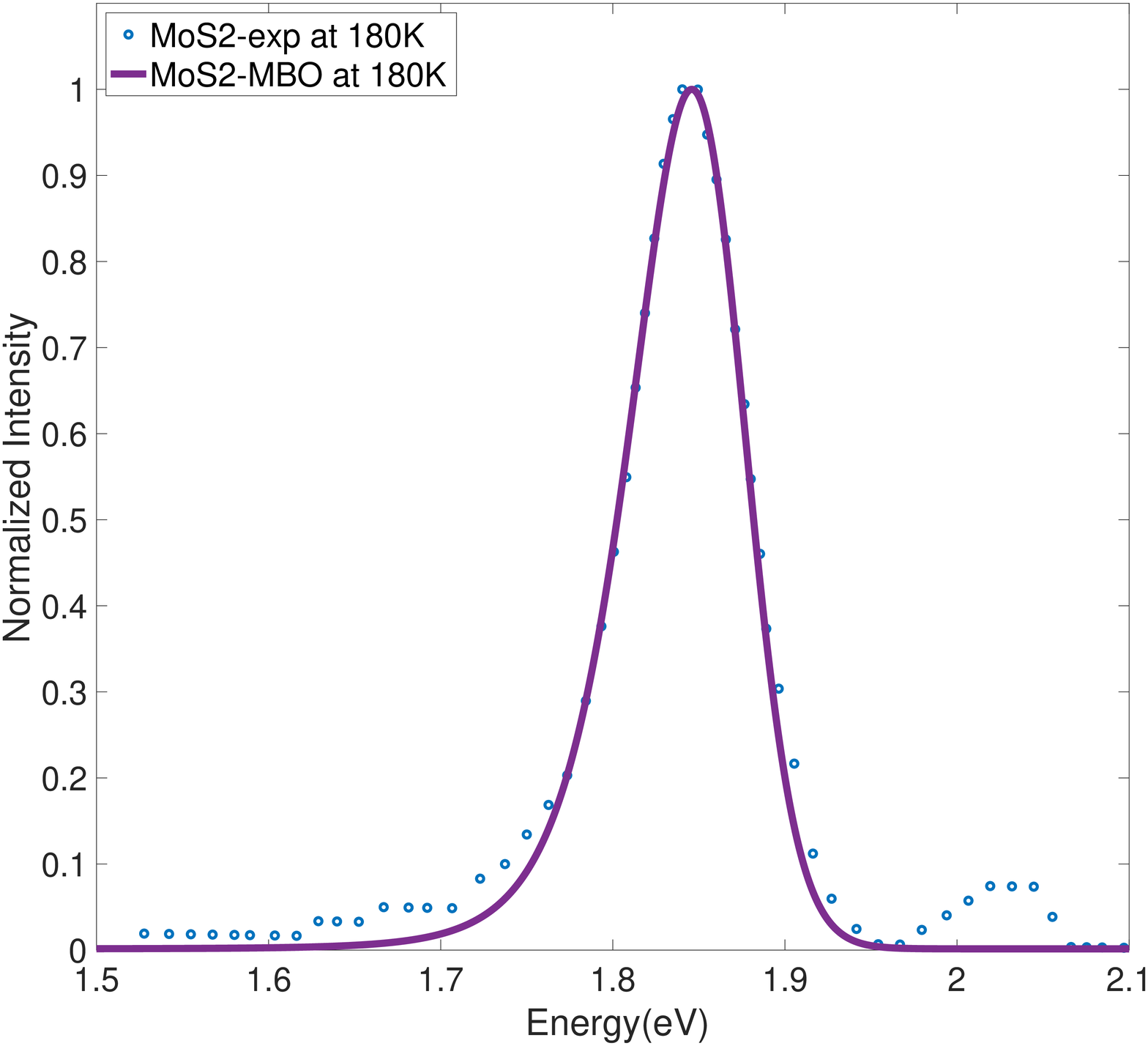}
}
\quad
\subfigure[]{
\includegraphics[scale=0.16,trim=100 0 40 0]{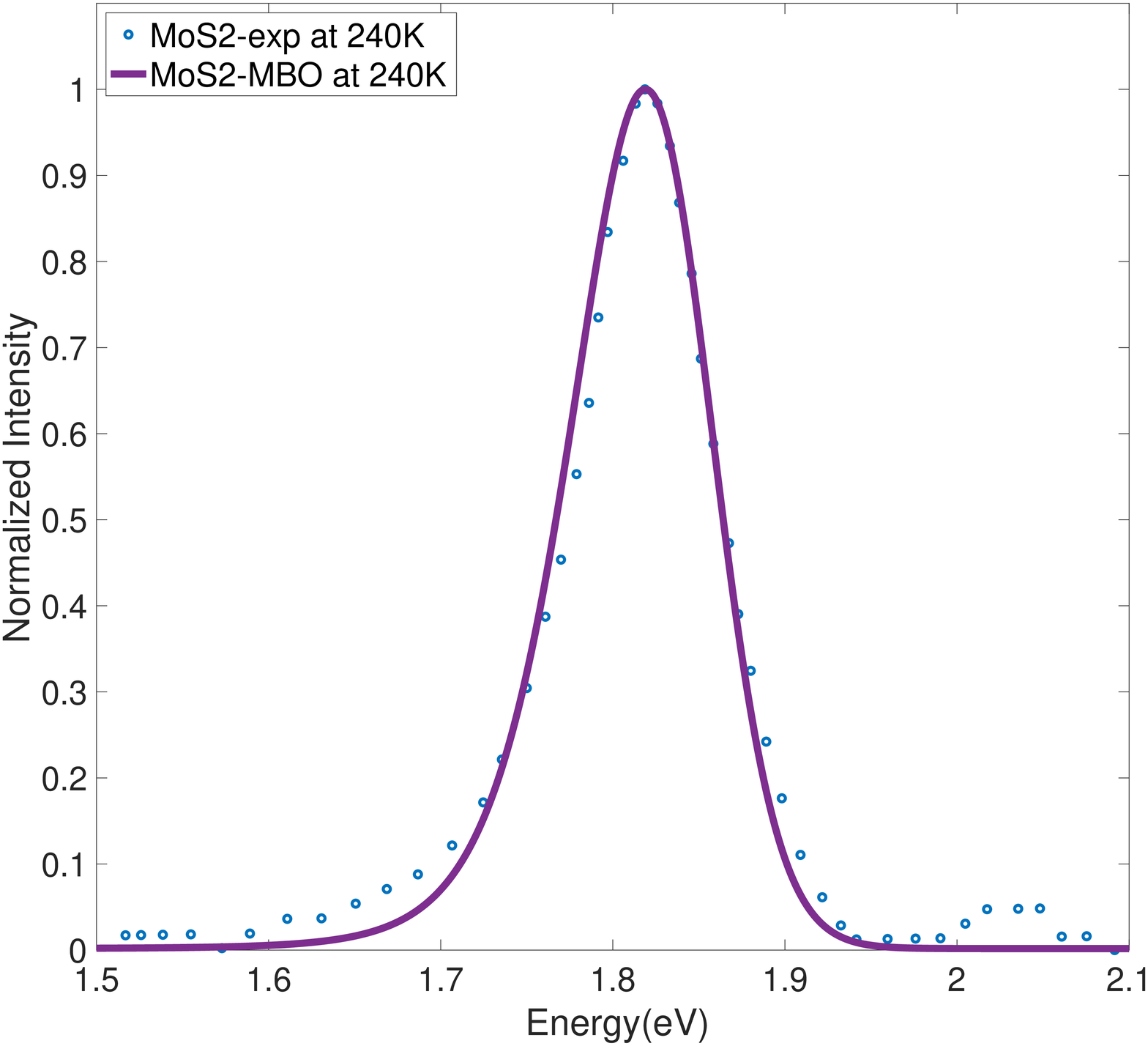}
}
\quad
\subfigure[]{
\includegraphics[scale=0.16,trim=100 0 40 0]{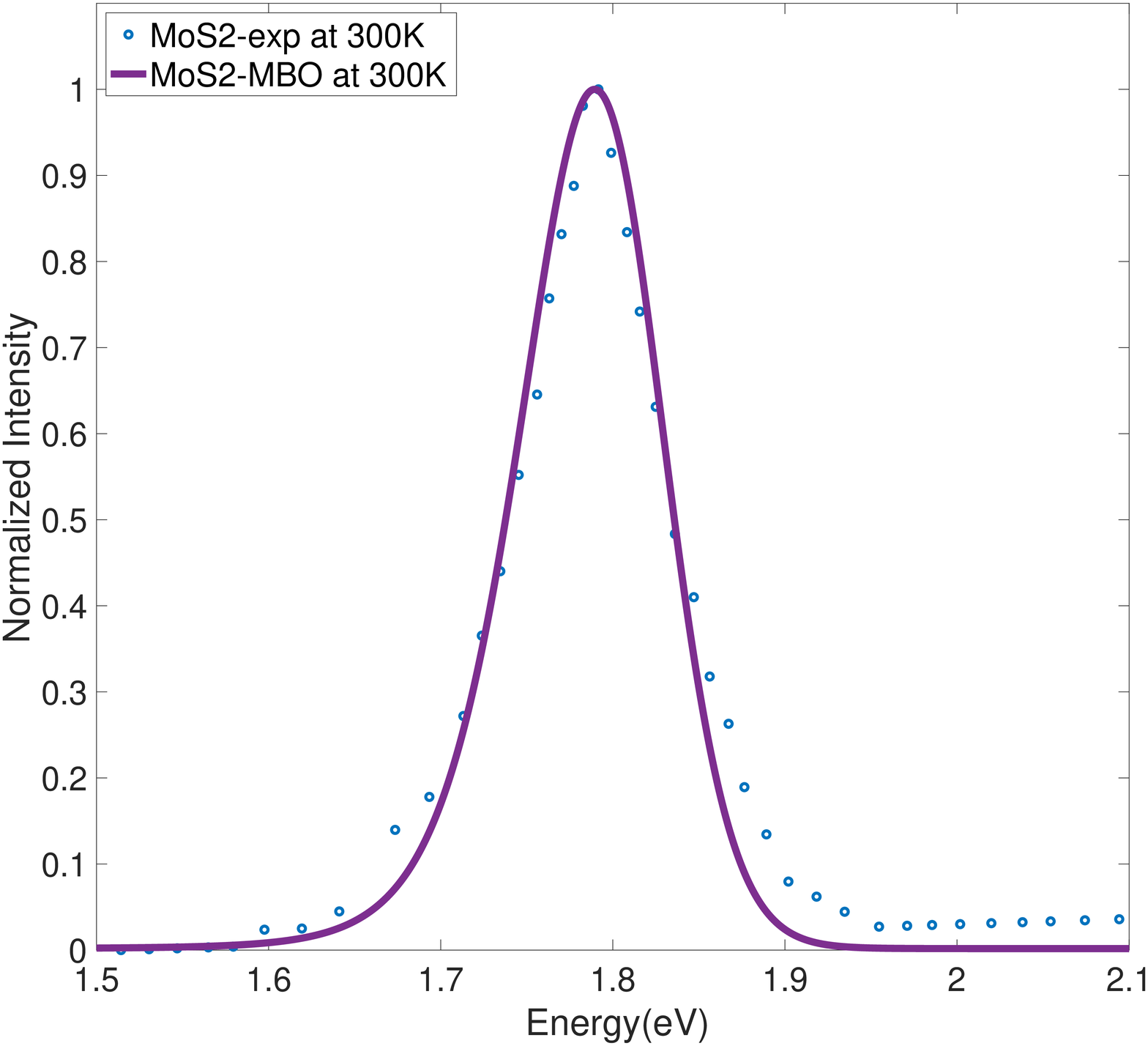}
}\\
\caption{Measured and MBO-fitted emission spectra of MoS$_2$ at different temperatures: (a) 4 K, (b) 30 K, (c) 60 K, (d) 90 K, (e) 120 K, (f) 150 K, (g) 180 K, (h) 240 K and (i) 300 K. MBO fitting parameters are collected in Table \ref{table4}.}
\label{fig4}
\end{figure*}

{Measured emission spectra~\cite{ref25} of a single-layer MoS$_2$ and MBO simulation are shown in Fig. \ref{fig4}}. With the same procedure, the major phonon frequency adopted from the Raman spectra~\cite{ref25} is about 385 cm$^{-1}$ (0.0477 eV). There are two pronounced peaks: the broad peak lying in low-energy domain is attributed to bound exciton state~\cite{spl, mak, coe}, which is absent in PL spectra of MoSe$_2$, and the narrow peak located in high-frequency scope is caused by free exciton emission. A small swelling or plateau is noticed near the 2.03 eV, which is given rise by the substrate~\cite{ref25}, a silicon wafer. Our MBO model reproduces the emission spectra with all necessary spectral features. The related simulation parameters are listed in the Table \ref{table4}, {where $\hbar\omega_{eg1}$ and $\hbar\omega_{eg2}$ are the ZPL energies for the bound exciton and free exciton states, respectively.} The underlying physics is revealed clearly by the decomposition of PL spectra into two kinds of exciton states with MBO model. As temperatures goes up, the peaks of bound exciton and free exciton are approaching each other in the spectra manifestation, meaning bound excitons obtain more thermal energy and the transition to free excitons occurs more easily, thus the energy gap between these two exciton states narrows. As for the contribution percentages $\alpha_1$ and $\alpha_2$ when superposing the above emission, the weighting of bound exciton emission gradually shifts to the emission of free excitation. Given the temperature in the range between 180 K and 300 K in Fig.~ \ref{fig4}(\rm{g})-(\rm{i}), MBO model can repeat the PL spectra of MoS$_2$ with only free exciton emission.

\begin{table*}[tbp]
\centering
\topcaption{Fitted MBO parameters for the PL spectra of MoS$_2$ at various temperatures. The units of $\omega$ and $\gamma$ are cm$^{-1}$, and that of $\hbar\omega_{eg}$ and T are eV and K, separately.}
\label{table4}
\begin{tabular}{| c | c | c | c | c | c | c | c | c | c | c |}
\hline
$\omega$ & $\gamma$ & $S_1$ & $S_2$ & $\hbar\omega_{eg1}$ & $\hbar\omega_{eg2}$ & $\hbar\omega_{eg2}-\hbar\omega_{eg1}$ & $\alpha_1(\%)$ & $\alpha_2(\%)$ & T \\\hline
385 & 1210 & 2.10 & 1.30 & 1.855 & 1.920 & 0.065 & 47.64 & 52.36 & 4 \\\hline
385& 1210 & 2.30 & 1.00 & 1.853 & 1.900 & 0.047 & 47.09 & 52.91 & 30 \\\hline
385 & 1210 & 2.30 & 0.80 & 1.855 & 1.895 & 0.040 & 44.44 & 55.56 & 60 \\\hline
385 & 1210 & 2.30 & 0.80 & 1.853 & 1.897 & 0.044 & 31.97 & 68.03 & 90 \\\hline
385 & 1210 & 2.30 & 0.80 & 1.855 & 1.885 & 0.030 & 21.26 & 78.74 & 120 \\\hline
385 & 1210 & 2.30 & 0.80 & 1.866 & 1.892 & 0.026 & 18.70 & 81.30 & 150 \\\hline
385 & 1210 & $\backslash$ & 0.85 & $\backslash$ & 1.876 & $\backslash$ & 0 & 100 & 180 \\\hline
385 & 1210 & $\backslash$ & 0.95 & $\backslash$ & 1.855 & $\backslash$ & 0 & 100 &240 \\\hline
385 & 1210 & $\backslash$ & 0.80 & $\backslash$ & 1.823 & $\backslash$ & 0 & 100 & 300 \\\hline
\end{tabular}
\end{table*}

\subsection{the modified Varshni equation}
\label{var}

Temperature-dependent positions of exciton peaks have been found in the PL of WSe$_2$,  WS$_2$, MoSe$_2$ and MoS$_2$, which can be attributed to temperature-induced lattice dilation and electron-phonon interactions. The Varshni equation~\cite{var}, known to be a great fit for temperature-dependent energy gaps for various semiconductors including TMDs, can be written as
\begin{align}
\hbar \omega_{eg}(T) =\hbar \omega_{eg}(0)-\alpha_{\rm v} T^2 /(T+\beta_{\rm D})
\end{align}
where $\omega_{eg}(T)$ is the energy of exciton peak at temperature $T$. $\alpha_{\rm v}$ and $\beta_{\rm D}$ are material constants with $\beta_{\rm D}$ related to the Debye temperature.

As an improvement to the Varshni equation incorporating additional physical information, a semi-empirical function $\hbar \omega_{eg}(T)$ has been widely applied in TMDs~\cite{hel,chr,raj,liz,rud}:
\begin{align}
\label{modi}
\hbar \omega_{eg}(T) = \hbar \omega_{eg}(0)-S_{\rm se}\langle\hbar \omega\rangle (\coth\frac{\langle\hbar \omega\rangle} {2k_B T}-1 )
\end{align}
Here $S_{\rm se}$ is the dimensionless interaction constant representing the intensity of exciton-phonon coupling, and $\langle\hbar \omega\rangle$ is the average phonon energy. Despite fitting deviations at low temperatures that are likely influenced by quantum effects, this semi-empirical fitting function works well at high temperatures. Therefore, we select high temperatures for WSe$_2$,  WS$_2$, MoSe$_2$ and MoS$_2$ to fit the equation, where the fitting HR factors tend to be constant. In this work, we use $S_{\rm MBO}$ and $S_{\rm se}$ to represent the exciton-phonon coupling strengths in MBO model and Eq.~(\ref{modi}), respectively.
{Although both of them are related to the measurement of the exciton-phonon interaction strengths, there are clear distinctions between the HR factor $S_{\rm MBO}$ and the empirical parameter $S_{\rm se}$. For $S_{\rm se}$, it is from the semi-empirical formula based on the previous spectra fitting~\cite{raj,liz,rud} while $S_{\rm MBO}$ is derived from the MBO model as follow:
\begin{align}
S_{\rm MBO} = \frac{m_j \omega_j d_j^2}{2 \hbar^2}
\end{align}
}
Thus, a dimensionless modifying factor $k$ is adopted to connect the $S_{\rm MBO}$ in our MBO model and $S_{se}$ in Eq.~(\ref{modi})
\begin{align}
S_{\rm MBO} = kS_{\rm se}
\end{align}

The connection of exciton-phonon coupling strength between MBO model and semi-empirical equation are listed in Table \ref{table5}, where $\omega$ is the primary phonon frequency and $\hbar \omega_{eg}$ is the ZPL energy. For WSe$_2$, we only consider the impact of phonon frequency 250 cm$^{-1}$ and ignore the phonon frequency 100 cm$^{-1}$ because its HR factors are too small. As for the remaining WS$_2$, MoSe$_2$ and MoS$_2$, the exciton emission separated by the MBO model from total PL line shape can be used for the fitting. Table \ref{table5} proves that a modifying factor exists in the range between 0.07 and 0.22, which should be a function of different TMDs, linking the MBO model and semi-empirical function.

\begin{table}
\centering
\topcaption{Fitted parameters of MBO and semi-empirical equation for the PL spectra of investigated TMDs at selected high temperatures. The units of $\omega$ is cm$^{-1}$, and those of $\hbar\omega_{eg}$ and T are eV and K, separately.}
\label{table5}
\begin{tabular}{| c | c | c | c | c | c | c | c |}
\hline
TMDs & $\omega$ & T & $\hbar\omega_{eg}$ & $S$ & $S_{\rm MBO}$ & $S_{\rm se}$ & $k$ \\\hline
\multirow{3}*{WSe$_2$} & \multirow{3}*{250} & 160 & 1.705 & 0.35 & \multirow{3}*{$0.33$} & \multirow{3}*{$2.304$} &  \multirow{3}*{$0.14$}\\
\cline{3-5}
~ & ~ & 240 & 1.678 & 0.30 & ~ & ~ & ~ \\
\cline{3-5}
~& ~ & 320 & 1.652 & 0.33 & ~ & ~ & ~ \\
\hline
\multirow{3}*{WS$_2$} & \multirow{3}*{352} & 180 & 2.053 & 0.20 & \multirow{3}*{$0.20$} & \multirow{3}*{$2.004$} &  \multirow{3}*{$0.10$}\\
\cline{3-5}
~ & ~ & 240 & 2.040 & 0.20 & ~ & ~ & ~ \\
\cline{3-5}
~& ~ & 295 & 2.026 & 0.20 & ~ & ~ & ~ \\
\hline
\multirow{4}*{MoSe$_2$} & \multirow{4}*{240} & 150 & 1.636 & 0.12 & \multirow{4}*{$0.18$} & \multirow{4}*{$2.591$} &  \multirow{4}*{$0.07$}\\
\cline{3-5}
~ & ~ & 190 & 1.614 & 0.15 & ~ & ~ & ~ \\
\cline{3-5}
~& ~ & 230 & 1.614 & 0.18 & ~ & ~ & ~ \\
\cline{3-5}
~& ~ & 270 & 1.589 & 0.25 & ~ & ~ & ~ \\
\hline
\multirow{3}*{MoS$_2$} & \multirow{3}*{385} & 180 & 1.876 & 0.85 & \multirow{3}*{$0.87$} & \multirow{3}*{$4.023$} &  \multirow{3}*{$0.22$}\\
\cline{3-5}
~ & ~ & 240 & 1.855 & 0.95 & ~ & ~ & ~ \\
\cline{3-5}
~& ~ & 300 & 1.823 & 0.80 & ~ & ~ & ~ \\
\hline
\end{tabular}
\end{table}

\section{Conclusions}
\label{Con}
The MBO model has been applied to model the {emission spectra} of TMD monolayers for the first time. Good agreement has been achieved between measured PL spectra of WSe$_2$,  WS$_2$, MoSe$_2$ and MoS$_2$, and their MBO fittings. Materials parameters extracted from the MBO fitting shed light on the interesting photophysics of various TMD monolayers with differing emission mechanisms, some of which involve bound exciton states and trions. For example, in our MBO investigation of WS$_2$, hybrid states of bound excitons and biexcitons, which degrade gradually and can be ignored at high temperatures, have considerable contributions to the emission spectra at low temperatures, such as 4 K and 30 K. However, the emission spectra of MoSe$_2$ are dominated by the trion and free exciton states at low temperatures. Moreover, those quasiparticles can also interact with each other and couple with photons, resulting in a robust exciton-trion polaritons~\cite{rana}.
For WSe$_2$,  WS$_2$ and MoSe$_2$, spectrally relevant vibrational modes with frequencies close to 30 meV are found to couple to electronic degrees of freedom at room temperature with corresponding HR factors of 0.3, 0.25 and 0.15, respectively. For MoS$_2$, the MBO model predicts a much larger frequency of 47.7 meV for the spectrally relevant vibrational mode, with a greater HR factor of 0.8. Another common spectral feature for WSe$_2$, WS$_2$ and MoS$_2$ is that there exists a certain temperature above which $S$ drops precipitously to a much lower value that no longer changes with temperature. A smaller S at higher temperatures can be attributed to the decrease in normal coordinate displacement caused by an increase of disorder-induced localization~\cite{Qi, mei,guha, hag}. Furthermore, the low HR factor is also related to geometric modifications such as the increase of bond length and bond angle~\cite{kaze} caused by high temperature.
Moreover, a modifying factor has been calculated to bridge fitted exciton-phonon coupling strengths to the semi-empirical Varshni equation for the first time. With its simplicity, the MBO model has been demonstrated as a valid approach for simulating linear optical spectra of TMD monolayers, opening up a new venue to understand optical properties of complex systems such as stacked heterostructures of TMDs.

{Despite that the MBO model is a successful phenomenological approach to optical spectra of TMD monolayers, a more sophisticated microscopic model describing simultaneous dynamics of interacting excitons, photons, and phonons will be highly beneficial to address exciton thermalization and diffusion, and to probe at a deeper level the dynamic, optical and transport properties of TMDs, such as the mechanism of PL of momentum-dark excitons\cite{bre}. For example, by applying the Heisenberg equation of motion with cluster expansion scheme, a quantum model including combined effects of coupled excitons, photons and phonons as well as consistent many-particle dephasing has been developed in Ref.~\cite{bre}.
{A potentially more advantageous alternative, the numerically exact approach employing the multi-$\rm D_2$ Davydov Ansatz~\cite{SKW,ZY} and the time-dependent variational principle have been successfully applied to various systems with combined light-matter-phonon interactions. It is our hope that the multi-$\rm D_2$ Ansatz can help unveil detailed exciton-photon-phonon dynamics and various spectroscopic signals in TMDs.
Specifically, if the excitation density is far from the exciton-Mott transition, a standard excitonic Hamiltonian in the rotating frame can be established, which neglects the fermionic substructure of excitons as well as exciton-exciton interactions. In this regime, the excitons can scatter from the valley $\rm K^{\prime}$ to $\rm K$ by virtue of emitting or absorbing a phonon, which is described by off-diagonal exciton-phonon coupling in the momentum state space. Thus, by constructing a linear superposition wave function of direct product states consisting of exciton momentum states and photon/phonon coherent states in the framework for the multi-$\rm D_2$ Ansatz, we can investigate the indirect PL signal of a momentum-forbidden dark exciton generating through phonon scattering. To account for the many-particle dephasing process, a continuum spectral density of a bath may have to be adopted, which has been achieved by the discretization scheme employed in the method of the multi-$\rm D_2$ Ansatz~\cite{SKW1}. Moreover, combining with the method of thermo-field dynamics, or the method of displaced number states, the multi-$\rm D_2$ Ansatz is easily extended to probe the temperature dependence of TMD optical properties.
Work in these directions are in progress.

\begin{acknowledgement}

The authors gratefully acknowledge the support of the Singapore Ministry of Education Academic Research Fund (Grant Nos. 2018-T1-002-175 and
2020-T1-002-075)). K.~Sun would also like to thank the Natural Science Foundation of Zhejiang Province (Grant No.~LY18A040005) for partial support.

\end{acknowledgement}

\section{Data Availability}
The data that support the findings of this study are available from the corresponding author upon reasonable request.



\begin{thebibliography}{999}

\bibitem{rad} Radisavljevic, B.; Radenovic, A .; Brivio, J.; Giacometti, V.; Kis, A. Single-layer MoS2 transistors. \emph{Nature. Nanotech}. \textbf{2011}, \emph{6}, 147–150.
\bibitem{mak2} Mak K. F.; He, K.; Lee, C.; Lee, G. H.; Hone, J.; Heinz, T. F.; Shan, J. Tightly bound trions in monolayer MoS2. \emph{Nature Mater}. \textbf{2013}, \emph{12}, 207–211.
\bibitem{spl} Splendiani, A.; Sun, L.; Zhang, Y.; Li, T.; Kim, J.; Chim, C.-Y.; Galli, G.; Wang, F. Emerging photoluminescence in monolayer MoS2. \emph{Nano Letters}. \textbf{2010}, \emph{10}, 1271–1275.
\bibitem{yan} Yang, L.; Xie, C.; Jin, J.; Ali, R. N.; Feng, C.; Liu, P.; Xiang, B. Properties, preparation and applications of low dimensional transition metal dichalcogenides. \emph{Nanomaterials}. \textbf{2018}, \emph{8}, 463.
\bibitem{di} Xiao, D.; Liu, G.-B.; Feng, W.; Xu, X.; Yao, W. Coupled spin and valley physics in monolayers of MoS2 and other group-VI dichalcogenides. \emph{Phys. Rev. Lett}. \textbf{2012}, \emph{108}, 196802.
\bibitem{kim} Kim, J.; Hong, X.; Jin, C.; Shi, S.-F.; Chang, C.-Y. S.; Chiu, M.-H.; Li, L.-J.; Wang, F. Ultrafast generation of pseudo-magnetic field for valley excitons in WSe2 monolayers. \emph{Science}. \textbf{2014}, \emph{346}, 1205–1208.
\bibitem{jones} Jones, A. M.; Yu, H.; Ghimire, N. J.; Wu, S.; Aivazian, G.; Ross, J. S.; Zhao, B.; Yan, J.; Mandrus, D. G.; Xiao, D.; Yao, W.; Xu, X. Optical generation of excitonic valley coherence in monolayer WSe2. \emph{Nature. Nanotech}. \textbf{2013}, \emph{8}, 634–638.
\bibitem{ros} Ross, J. S.; Wu, S.; Yu, H.; Ghimire, N. J.; Jones, A. M.; Aivazian, G.; Yan, J.; Mandrus, D. G.; Xiao, D.; Yao, W.; Xu, X. Electrical Control of Neutral and Charged Excitons in a Monolayer Semiconductor. \emph{Nat. Commun}. \textbf{2013}, \emph{4}, 1474.
\bibitem{fink} Finkelstein, G.; Shtrikman, H.; Bar-Joseph, I. Optical spectroscopy of a two-dimensional electron gas near the metal-insulator transition. \emph{Phys. Rev. Lett}. \textbf{1995}, \emph{74}, 976.
\bibitem{hua} Huard, V.; Cox, R. T.; Saminadayar, K.; Arnoult, A.; Tatarenko, S. Bound states in optical absorption of semiconductor quantum wells containing a two-dimensional electron gas. \emph{Phys. Rev. Lett}. \textbf{2000}, \emph{84},187.
\bibitem{muk} Mukamel, S. \emph{Principles of nonlinear optical spectroscopy}, Oxford University Press on Demand, \textbf{1999}.
\bibitem{knox} Zhao, Y; Knox, R. S. A brownian oscillator approach to the Kennard-Stepanov relation. \emph{J. Phys. Chem. A}. \textbf{2000}, \emph{104}, 7751-7761.
\bibitem{xu} Xu, S. J.; Li, G. Q.; Wang, Y. J.; Zhao, Y.; Chen, G. H.; Zhao, D. G.; Zhu, J. J; Yang, H.; Yu, D. P.; Wang, J. N. Quantum dissipation and broadening mechanisms due to electron-phonon interactions in self-formed InGaN quantum dots. \emph{Applied physics letters}. \textbf{2006}, \emph{88}, 083123.
\bibitem{huang} Huang, K.; Rhys, A. \emph{Selected Papers Of Kun Huang:(With Commentary), World Scientific}. \textbf{2000}, pp. 74–92.
\bibitem{ref14} Ye, J.; Grimsdale, A. C.; Zhao, Y. Analyzing the optical properties of a conjugated polymer by the multimode Brownian oscillator model. \emph{J. Phys. Chem. A}. \textbf{2010}, \emph{114}, 504–508.
\bibitem{ref15} Sun, K.; Liu, X.; Hu, W.; Zhang, M.; Long, G.; Zhao, Y. Singlet fission dynamics and optical spectra of pentacene and its derivatives. \emph{Phys. Chem. Chem. Phys}. \textbf{2021}, \emph{23}, 12654–12667.
\bibitem{Qi} Qi, P.; Luo, Y.; Shi, B.; Li, W.; Liu, D.; Zheng, L.; Liu, Z.; Hou, Y.; Fang, Z. Phonon scattering and exciton localization: molding exciton flux in two dimensional disorder energy landscape. \emph{eLight}. \textbf{2021}, \emph{1}, 6.
\bibitem{chr} Christiansen, D.; Selig, M.; Berghäuser, G.; Schmidt, R.; Niehues, I.; Schneider, R.; Arora, A.; de Vasconcellos, S. M.; Bratschitsch, R.; Malic, E.; Knorr, A. Phonon sidebands in monolayer transition metal dichalcogenides. \emph{Phys. Rev. Lett}. \textbf{2017}, \emph{119}, 187402.
\bibitem{bre} Brem, S.; Ekman, A.; Christiansen, D.; Katsch, F.; Selig, M.; Robert, C.; Marie, X.; Urbaszek, B.; Knorr, A.; Malic, E. Phonon-assisted photoluminescence from indirect excitons in monolayers of transition-metal dichalcogenides. \emph{Nano Lett}. \textbf{2020}, \emph{20}, 2849-2856.
\bibitem{dam} Zhao, Y.; Chernyak, V.; Mukamel, S. Spin versus boson baths in nonlinear spectroscopy. \emph{J. Phys. Chem. A}. \textbf{1998}, \emph{102}, 6614–6634.
\bibitem{selig} Selig, M.; Berghäuser, G.; Richter, M.; Bratschitsch, R.; Knorr, A.; Malic, E. Dark and bright exciton formation, thermalization, and photoluminescence in monolayer transition metal dichalcogenides. \emph{2D Materials}. \textbf{2018}, \emph{5}, 035017.
\bibitem{zhao} Zhao, H.; Kalt, H. Energy-dependent Huang-Rhys factor of free excitons. \emph{Phys. Rev. B}. \textbf{2003}, \emph{68}, 125309.
\bibitem{miti} Mitioglu, A. A.; Plochocka, P.; Jadczak, J. N.; Escoffier, W.; Rikken, G. L. J. A.; Kulyuk, L.; Maude, D. K. Optical manipulation of the exciton charge state in single-layer tungsten disulfide. \emph{Phys. Rev. B}. \textbf{2013}, \emph{88}, 245403.
\bibitem{zhu} Zhu, B.; Chen, X.; Cui, X. Exciton binding energy of monolayer WS2. \emph{Sci Rep}, \textbf{2015}, \emph{5}, 9218.
\bibitem{pei} Peimyoo, N.; Yang, W.; Shang, J.; Shen, X.; Wang, Y.; Yu, T. Chemically driven tunable light emission of charged and neutral excitons in Monolayer WS2. \emph{ACS Nano}. \textbf{2014}, \emph{8}, 11320–11329.
\bibitem{bel} Bellus, M. Z.; Ceballos, F.; Chiu, H. Y.; Zhao, H. Tightly bound trions in transition metal dichalcogenide heterostructures. \emph{ACS Nano}, \textbf{2015}, \emph{9}, 6459–6464.
\bibitem{ref17} Plechinger, G.; Nagler, P.; Kraus, J.; Paradiso, N.; Strunk, C.; Schüller, C.; Korn, T. Identification of excitons, trions and biexcitons in single-Layer WS2. \emph{Physica Status Solidi - Rapid Research Letters}. \textbf{2015}, \emph{9}, 457–461.
\bibitem{ref18} Zeng, H.; Liu, G. bin; Dai, J.; Yan, Y.; Zhu, B.; He, R.; Xie, L.; Xu, S.; Chen, X.; Yao, W.; Cui, X. Optical signature of symmetry variations and spin-valley coupling in atomically thin tungsten dichalcogenides. \emph{Sci Rep}. \textbf{2013}, \emph{3}, 1608.
\bibitem{ref19} Chernikov, A.; Berkelbach, T. C.; Hill, H. M.; Rigosi, A.; Li, Y.; Aslan, O. B.; Reichman, D. R.; Hybertsen, M. S.; Heinz, T. F. Phys. Exciton binding energy and nonhydrogenic Rydberg Series in monolayer WS2. \emph{Phys. Rev. Lett}. \textbf{2014},
\emph{113}, 076802.
 \bibitem{ref20} Ye, Z.; Cao, T.; O’Brien, K.; Zhu, H.; Yin, X.; Wang, Y.; Louie, S. G.; Zhang, X. Probing excitonic dark states in single-layer tungsten disulphide. \emph{Nature}. \textbf{2014}, \emph{513}, 214–218.
\bibitem{ref21} Zhu, B.; Chen, X.; Cui, X. Exciton binding energy of monolayer WS2. \emph{Sci. Rep.} \textbf{2015}, \emph{5}, 9218.
\bibitem{ref22} Hanbicki, A. T.; Currie, M.; Kioseoglou, G.; Friedman, A. L.; Jonker, B. T. Measurement of high exciton binding energy in the monolayer transition-metal dichalcogenides WS2 and WSe2. \emph{Solid State Commun}. \textbf{2015}, \emph{203}, 16–20.
\bibitem{ref23} Tonndorf, P.; Schmidt, R.; Böttger, P.; Zhang, X.; Börner, J.; Liebig, A.; Albrecht, M.; Kloc, C.; Gordan, O.; Zahn, D. R. T.; Michaelis de Vasconcellos, S.; Bratschitsch, R. Photoluminescence emission and Raman response of monolayer MoS2, MoSe2, and WSe2. \emph{Optics Express}. \textbf{2013}, \emph{21}, 4908.
\bibitem{ref24} Tongay,S.; Zhou, J.; Ataca, C.; Lo, K.; Matthews, T. S.; Li, J.; Grossman, J. C.; Wu, J. Thermally driven crossover from indirect toward direct bandgap in 2D semiconductors: MoSe2 versus MoS2. \emph{Nano Letters}. \textbf{2012}, \emph{12}, 5576–5580.
\bibitem{mak1} Mak, K. F.; He, K.; Shan, J.; Heinz, T. F. Control of valley polarization in monolayer MoS2 by optical helicity. \emph{Nat. Nanotech}. \textbf{2012}, \emph{7}, 494–498.
\bibitem{evan} Evans, B. L.; Hazelwood, R. A. Optical and structural properties of MoSe2. \emph{Physica Status Solidi (a)}. \textbf{1971}, \emph{4}, 181–192.
\bibitem{bor1} Borrelli, R.; Peluso, A. The temperature dependence of radiationless transition rates from ab initio computations. \emph{Phys. Chem. Chem. Phys}. \textbf{2011}, \emph{13}, 4420-4426.
\bibitem{bor2} Borrelli, R.; Capobianco, A.; Peluso, A. Generating function approach to the calculation of spectral band shapes of free-base chlorin including Duschinsky and Herzberg–Teller effects. \emph{J. Phys. Chem. A}. \textbf{2012}, \emph{116}, 9934-9940.
\bibitem{bor3} Borrelli, R.; Ellena, S.; Barolo, C. Theoretical and experimental determination of the absorption and emission spectra of a prototypical indolenine-based squaraine dye. \emph{Phys. Chem. Chem. Phys}. \textbf{2014}, \emph{16}, 2390-2398.
\bibitem{capo} Capobianco, A.; Borrelli, R.; Landi, A.; Velardo, A.; Peluso, A. Absorption band shapes of a push–pull dye approaching the cyanine limit: A challenging case for first principle calculations. \emph{J. Phys. Chem. A}. \textbf{2016}, \emph{120}, 5581-5589.
\bibitem{ref25} Korn, T.; Heydrich, S.; Hirmer, M.; Schmutzler, J.; Schller, C. Low-temperature photocarrier dynamics in monolayer MoS2. \emph{Appl. Phys. Lett}. \textbf{2011}, \emph{99}, 102109.
\bibitem{mak} Mak, K.; Lee, C.; Hone, J.; Shan, J.; Heinz, T. Atomically thin MoS2: a new direct-gap semiconductor. \emph{Phys. Rev. Lett}. \textbf{2010}, \emph{105}, 2–5.
\bibitem{coe} Coehoorn, R.; Haas, C.; De Groot, R. Electronic structure of MoSe2, MoS2, and WSe2. II. The nature of the optical band gaps. \emph{Phys. Rev. B}. \textbf{1987}, \emph{35}, 6203–6206.
\bibitem{var} Varshni, Y. P. Temperature dependence of the energy gap in semiconductors. \emph{Physica}. \textbf{1967}, \emph{34}, 149-154.
\bibitem{hel} Helmrich, S.; Schneider, R.; Achtstein, A. W.;  Arora, A.; Herzog, B.; de Vasconcellos, S. M.; Kolarczik, M.; Schöps, O.; Bratschitsch, R.; Woggon, U.;  Owschimikow, N. Exciton–phonon coupling in mono- and bilayer MoTe2. \emph{2D Mater}. \textbf{2018}, \emph{5}, 045007.
\bibitem{raj} Raja, A.; Selig, M.; Berghäuser, G.; Yu, J.; Hill, H. M.; Rigosi, A. F.; Brus, L. E.; Knorr, A.; Heinz, T. F.; Malic, E.; Chernikov, A. Enhancement of Exciton–Phonon Scattering from Monolayer to Bilayer WS2. \emph{Nano Letters}. \textbf{2018}, \emph{18}, 6135–6143.
\bibitem{liz} Li, H.; Zhu, X.; Tang, Z.; Zhang, X. Low-temperature photoluminescence emission of monolayer MoS2 on diverse substrates grown by CVD. \emph{Journal of Luminescence}. \textbf{2018}, \emph{199}, 210–215.
\bibitem{rud} Rudin, S.; Reinecke, T. L.; Segall, B. Temperature-dependent exciton linewidths in semiconductors. \emph{Phys. Rev. B, Condensed Matter}. \textbf{1990}, \emph{42}, 11218–11231.
\bibitem{rana} Rana, F.; Koksal, O.; Jung, M.; Shvets, G.; Vamivakas, A. N.; Manolatou, C.  Exciton-trion polaritons in doped two-dimensional semiconductors. \emph{Phys. Rev. Lett}. \textbf{2021}, \emph{126}, 127402–127402.
\bibitem{mei} Meier, T.; Zhao, Y.; Chernyak, V.; Mukamel, S. Polarons, localization, and excitonic coherence in superradiance of biological antenna complexes. \emph{J. Chem. Phys}. \textbf{1997}, \emph{107}, 3876–3893.
\bibitem{guha} Guha, S.; Rice, J. D.; Yau, Y. T.; Martin,C. M.; M Chandrasekhar; Chandrasekhar, H. R.; Guentner, R.; Scanduicci de Freitas, P.; Scherf, U. Temperature-dependent photoluminescence of organic semiconductors with varying backbone conformation. \emph{Phys. Rev. B}. \textbf{2003}, \emph{67}, 125204.
\bibitem{hag} Hagler, T. W.; Pakbaz, K.; Voss, K. F.; Heeger, A. J. Enhanced order and electronic delocalization in conjugated polymers oriented by gel processing in polyethylene. \emph{Phys. Rev. B}. \textbf{1991}, \emph{44}, 8652.
\bibitem{kaze} Kazemi, S. A.; Yengejeh, S. I.; Wang, V.; Wen, W.; Wang, Y. Theoretical understanding of electronic and mechanical properties of 1T' transition metal dichalcogenide crystals. \emph{Beilstein journal of nanotechnology}. \textbf{2022}, \emph{13}, 160-171.

\bibitem{ZY} Zhao, Y.; Sun, K.; Chen, L.; Gelin, M. F. The hierarchy of Davydov’s Ansätze and its applications. \emph{Wiley Interdisciplinary Reviews: Computational Molecular Science.} \textbf{2021}, \emph{e1589}.
\bibitem{SKW} Sun, K.; Dou, C.; Gelin, M. F.; Zhao, Y. Dynamics of disordered Tavis–Cummings and Holstein–Tavis–Cummings models. \emph{J. Chem. Phys.} \textbf{2022}, \emph{156}, 024102.
\bibitem{SKW1} Sun, K.; Huang, Z.; Gelin, M. F.; Chen, L.; Zhao, Y. Monitoring of singlet fission via two-dimensional photon-echo and transient-absorption spectroscopy: Simulations by multiple Davydov trial states. \emph{J. Chem. Phys.} \textbf{2019}, \emph{151}, 114102.





\end{thebibliography}

\end{document}